\documentclass[prl,amsmath,amssymb,aps,superscriptaddress,floatfix,reprint,notitlepage]{revtex4-1}
\usepackage{natbib}
\usepackage{graphicx}
\usepackage{color}
\usepackage{dcolumn}
\usepackage{bm}
\usepackage[utf8]{inputenc}
\usepackage{amssymb}
\usepackage{color,amsmath}
\usepackage{mathtools,ulem}
\usepackage{soul,xcolor} 
\setstcolor{red} 
\usepackage{epstopdf}
\usepackage{scalerel}
\usepackage{hyperref}
\usepackage{chemformula}
\usepackage{gensymb}
\usepackage{braket}
\hypersetup{
 pdfnewwindow=true, colorlinks=true,
 linkcolor=blue, anchorcolor=blue,
 citecolor=blue, filecolor=blue,
 menucolor=blue, urlcolor=blue}
\usepackage{array}

\begin{document}

\title{Engineering electrically-switchable quantum anomalous Hall states by spin-orbit coupling}

\author{Maosen Qin} 
\affiliation{Institute of Science and Technology Austria, Klosterneuburg, Austria, 3400}
\author{Ziwei Wang} 
\affiliation{Rudolf Peierls Centre for Theoretical Physics, Clarendon Laboratory, Parks Road, Oxford, OX1 3PU, UK}\author{Gyeongmin Kim} 
\affiliation{Institute of Science and Technology Austria, Klosterneuburg, Austria, 3400}
\author{Kenji Watanabe} 
\affiliation{Research Center for Electronic and Optical Materials, National Institute for Materials Science, 1-1 Namiki, Tsukuba 305-0044, Japan}
\author{Takashi Taniguchi} 
\affiliation{Research Center for Materials Nanoarchitectonics, National Institute for Materials Science,  1-1 Namiki, Tsukuba 305-0044, Japan}
\author{Steven H. Simon} 
\affiliation{Rudolf Peierls Centre for Theoretical Physics, Clarendon Laboratory, Parks Road, Oxford, OX1 3PU, UK}\author{Siddharth A. Parameswaran} 
\affiliation{Rudolf Peierls Centre for Theoretical Physics, Clarendon Laboratory, Parks Road, Oxford, OX1 3PU, UK}
\author{Hryhoriy Polshyn}   
\email{hpolshyn@ist.ac.at}
\affiliation{Institute of Science and Technology Austria, Klosterneuburg, Austria, 3400}

\maketitle
\textbf{Nonvolatile gate-driven switching of quantum anomalous Hall (QAH) states in graphene moir\'e systems~\cite{polshyn_electrical_2020, zhang_local_2023, zhang_manipulation_2024, stepanov_competing_2021, grover_chern_2022, su_moire-driven_2025, choi_superconductivity_2025, wang_family_2026, chen_layer-engineered_2026, wang_electrical_2025} provides a promising route toward topological electronics based on chiral edge states~\cite{chang_colloquium_2023}.
However, deliberate use of this switching mechanism requires control over both the magnetic properties and metastability of QAH states. 
While previous demonstrations mostly relied on the intrinsic magnetic energy landscape of moir\'e devices, here we show that this landscape can be engineered through proximity coupling to WSe\textsubscript{2}. 
We find that proximitizing twisted monolayer-bilayer graphene~\cite{polshyn_electrical_2020, chen_electrically_2021, he_competing_2021, xu_tunable_2021,  polshyn_topological_2022, waters_topological_2024,  zhang_manipulation_2024, zhang_local_2023, peng_abundant_2024} by WSe\textsubscript{2} reshapes the magnetization reversals responsible for nonvolatile electrical switching of QAH states.
We attribute this effect to the proximity-induced spin-orbit coupling (SOC), which can lock spin and valley and modify the magnetization of the competing states involved in switching compared with non-proximitized graphene systems. 
Our findings establish proximity-induced SOC as a new way to engineer magnetic properties and switchable magnetic states in graphene-based systems.
We further demonstrate that strong magnetic metastability in tMBG allows the magnetic states to be gate-tuned between QAH and metallic regimes, and between QAH states with Chern numbers $|C|=2$ and $1$ without resetting the magnetic state. 
This functionality points toward new device architectures based on QAH chiral edge states. 
} 

\begin{figure*}[ht!]
\includegraphics[width=\textwidth]{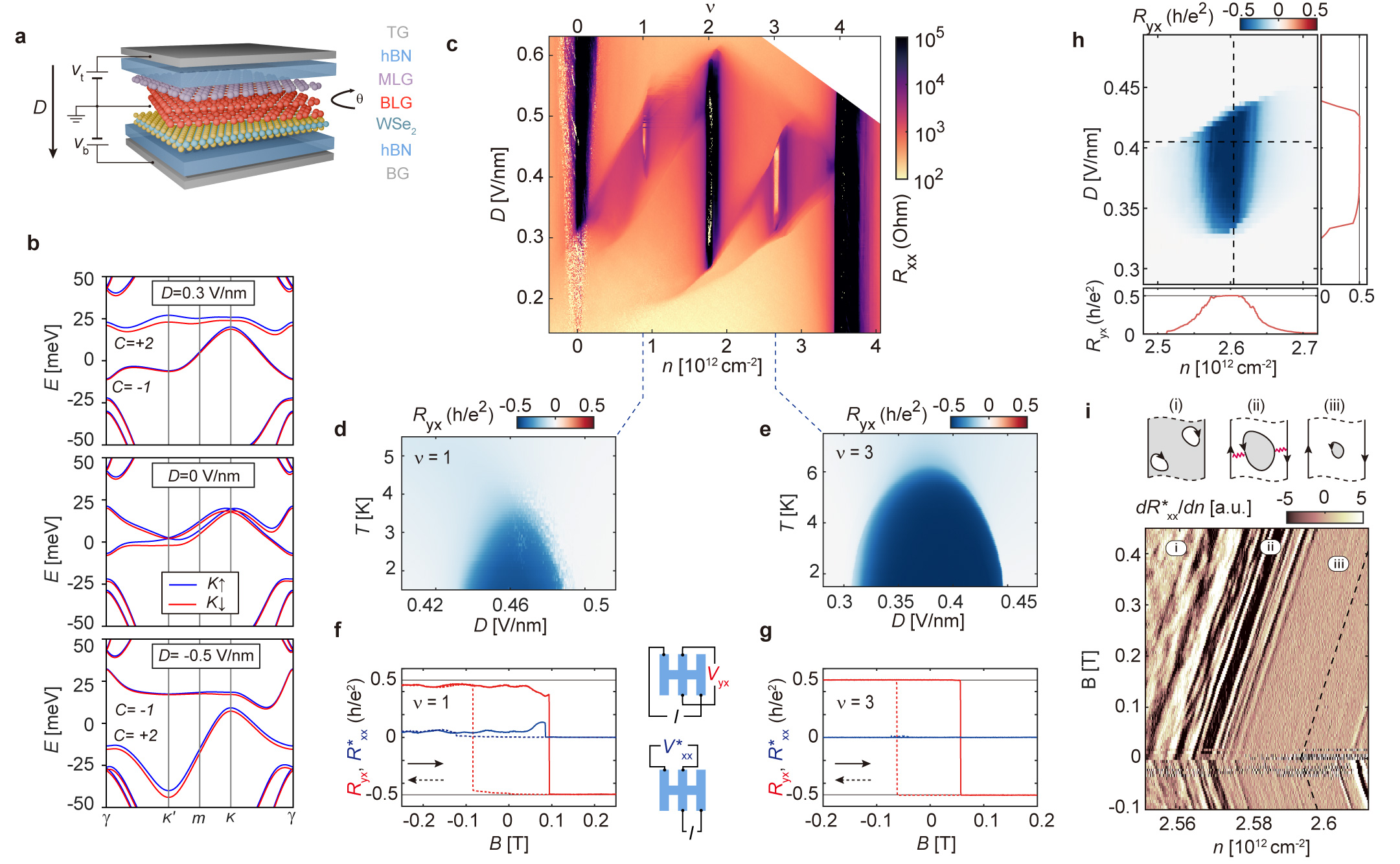} 
\caption{\textbf{Quantized anomalous Hall (QAH) effect in tMBG/WSe\textsubscript{2} device with $\theta= 1.23\degree$ (D2).}
\textbf{a,} Schematic of a tMBG/WSe\textsubscript{2} device consisting of rotationally misaligned monolayer graphene(MLG) and Bernal-stacked bilayer graphene(BLG) placed on a monolayer WSe\textsubscript{2}, encapsulated by hexagonal boron nitride (hBN).  
Top and bottom graphite gates control carrier density $n$ and displacement field $D$. 
\textbf{b,} Calculated band structure of tMBG/WSe\textsubscript{2} with proximity-induced Ising spin-orbit coupling of $\lambda_I=4$~meV at $D$ = 0.3, 0 and -0.5~V/nm.
\textbf{c,} Longitudinal resistance $R_{xx}$ measured in the conduction band at {50~mT} and $D>0$.
Sharp dips in $R_{xx}$ at $\nu=1$ and $3$ mark QAH states surrounded by higher-resistance metallic isospin-ferromagnetic regions. 
\textbf{d,  e,} Hall resistance $R_{yx}$ measured as a function of temperature $T$ and displacement field $D$ at {$B$ = 100~mT} for $\nu$= 1 (d) and $\nu$=3 (e), respectively. 
Dome-shaped regions with $R_{yx}\approx -h/2e^2$ mark stable $|C|=2$ QAH states.
\textbf{f,g,}~Magnetic-field dependence of $R_{yx}$ and $R^*_{xx}$ measured at $\nu=1$, {$D=0.48$V/nm}~(f) and $\nu=3$, {$D=0.42$V/nm}~(g) showing QAH hysteresis. 
\textbf{h,} Antisymmetrized Hall resistance $R_{yx}$ measured as a function of $n$ and $D$ near $\nu=3$ at $B=50$~mT, showing a quantized QAH plateau.
\textbf{i,} $dR^*_{xx}/dn$ measured as a function of $n$ and $D$ near $\nu=3$ at $B=50$~mT. 
The derivative highlights featureless regions of clean edge-state transport with decoupled edge states (iii) and bulk transport (i), and sharp peaks at plateau transitions due to defect-mediated edge-state coupling (ii), as schematically shown in the insets.
$dR^*_{xx}$ was measured by sweeping the bottom gate at fixed top-gate voltage, corresponding to $D\approx0.41$~V/nm.
}
\label{fig:1}
\end{figure*}

\begin{figure*}[ht!]
\includegraphics[width=\textwidth]{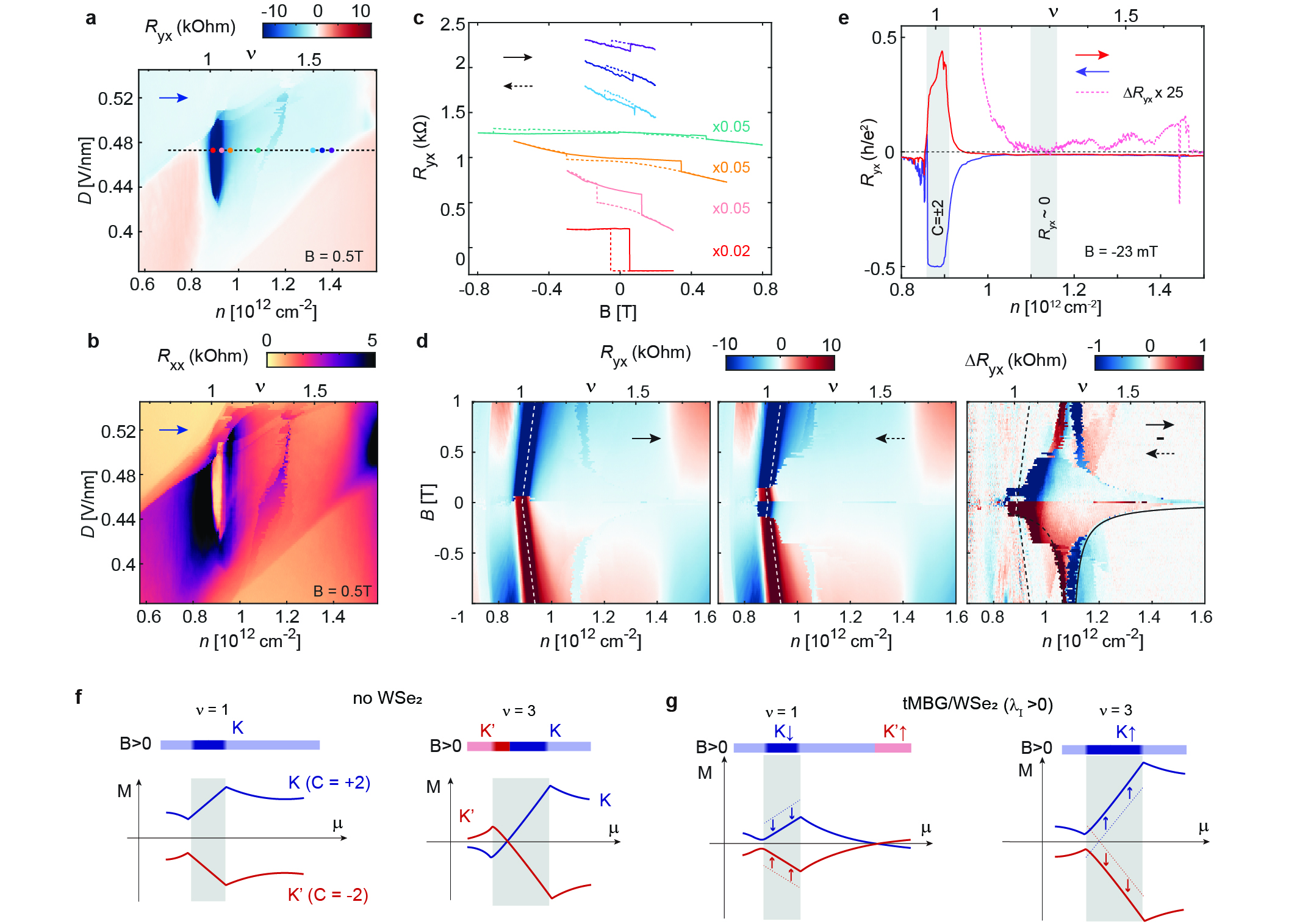} 
\caption{\textbf{Magnetization reversal and nonvolatile electrical switching near $\nu=1$ QAH state in device D2.}
\textbf{a,b,} $R_{yx}$ (a) and $R_{xx}$ (b) as a function of $n$ and $D$, measured at $B$= 0.5~T. 
The fast sweep axis is $n$, with the direction indicated by arrows.
In addition to QAH state at $\nu=1$, a sharp change of the resistance at intermediate fillings near $\nu\approx 1.2-1.3$ marks voltage-induced switching between magnetic states.  
\textbf{c,} Magnetic hysteresis of $R_{yx}$ measured at $D = 0.47$~V/nm and  $n$=1.40, 1.36, 1.32, 1.08, 0.96, 0.93, 0.89~$\times10^{12} \mathrm{cm}^{-2}$, as indicated in \textbf{a}. 
Curves are offset and scaled for clarity, as indicated.  
\textbf{d,} $R_{yx}$ measured as a function of $B$ at $D = 0.47$~V/nm for opposite $n$ sweep directions as indicated in the left and middle subpanels. 
The right subpanel shows the difference $\Delta R_{yx}$ between the two measurements.
Dashed straight lines mark the slopes expected for states with Chern numbers $C = \pm2$. 
Solid and dashed hyperbolic lines near the transition are guide to the eye following $ B \sim 1/(n - n_{rev})$, where $n_{rev}$ is the carrier density corresponding to the magnetization reversal point.
\textbf{e,} Nonvolatile gate-driven switching  of the $\nu=1$ QAH state measured at $B= -23$~mT. 
Red and blue traces correspond to the forward and backward scans of $n$, respectively. 
The dashed trace is a scaled difference, $\Delta R_{yx}\times 25$, between forward and backward sweeps.  
\textbf{f, g} Schematics showing the evolution of magnetization as a function of chemical potential for the two states involved in gate-driven switching. The magnetization reversal point at $\nu = 3$ in tMBG devices (f) is removed by spin-valley locking induced by Ising SOC (g).
 }
\label{fig:2}
\end{figure*}

In zero magnetic field, QAH states are bistable between two  magnetic states that have the opposite Chern numbers and opposite edge state chiralities.  The ease of electrical readout 
through the sign of the quantized Hall resistance~\cite{chang_colloquium_2023},  
and the absence of a strong magnetic field requirement make QAH states promising for applications in metrology~\cite{huang_quantum_2025}, low-power magnetic memory~\cite{alam_non-volatile_2021}, in-memory computing~\cite{liu_cryogenic_2025, alam_cryocim_2022}, microwave devices based on reconfigurable edge states~\cite{viola_hall_2014, martinez_circulators_2025}, and hybrid QAH-superconductor devices~\cite{clarke_exotic_2014,lian_topological_2018}.
Realizing these applications, however, requires efficient and nonvolatile control of QAH states, motivating recent efforts to manipulate these states electrically~\cite{serlin_intrinsic_2020, tschirhart_intrinsic_2023, yuan_electrical_2024} 
and  optically~\cite{huber_optical_2026, holtzmann_optical_2026, cai_optical_2026}.
Gate-voltage-driven, nonvolatile electrical control is particularly attractive because it could enable low-dissipation operation and dense electronic integration. 
It is also highly desirable to extend gate-control from mere switching of the sign of Chern number, to tuning between distinct Chern numbers or metallic phases, as this enables new operation modes for potential devices. 

The orbital character and high gate tunability of magnetism in graphene moir\'e systems~\cite{liu_orbital_2021} enable a distinct form of nonvolatile gate-driven switching of the magnetic and QAH states.   
This switching exploits regimes in which the magnetization changes sign as a function of chemical potential or displacement field, both of which can be controlled by electrical gating~\cite{zhu_voltage_2020}.
Nonvolatile switching is achieved by electrically tuning the system across such a magnetization reversal (MR), thereby reversing the stability of the magnetic states in small external magnetic field.
MRs can commonly occur within the QAH plateau when the chemical potential crosses the Chern gap, resulting in a rapid change and sign reversal of the orbital magnetization~\cite{zhu_voltage_2020}. 
Such behavior has been observed in twisted monolayer-bilayer graphene(tMBG)~\cite{polshyn_electrical_2020,  zhang_local_2023, zhang_manipulation_2024}, twisted bilayer graphene~\cite{stepanov_competing_2021, grover_chern_2022, lin_spin-orbitdriven_2022, tseng_anomalous_2022, bhowmik_spin-orbit_2023, su_moire-driven_2025}, and rhombohedral graphene multilayers~\cite{sha_observation_2024,  choi_superconductivity_2025, wang_family_2026, chen_layer-engineered_2026, wang_electrical_2025}. 
However, MRs can also occur in metallic regimes and even in systems without superlattices~\cite{han_orbital_2023}.
The ability to tune the magnetization and the metastability of QAH and related metallic magnetic states is crucial for the efficient control of QAH states. 
How these quantities can be purposefully engineered, however, remains largely unexplored in graphene systems.

Spin-orbit coupling (SOC) provides a natural route to controlling magnetic properties in two-dimensional systems because it couples spin and orbital degrees of freedom.  
Although intrinsic SOC in graphene is weak, proximity to transition metal dichalcogenides (TMDs) can induce SOC of up to several meV~\cite{wang_origin_2016, gmitra_proximity_2017, khoo_-demand_2017, island_spinorbit-driven_2019, wang_quantum_2019}. 
In graphene moir\'e systems, proximity to TMDs has previously been demonstrated to stabilize superconductivity~\cite{arora_superconductivity_2020, su_superconductivity_2023} and ferromagnetic states~\cite{lin_spin-orbitdriven_2022, bhowmik_spin-orbit_2023}.
Here, we show that proximity-induced SOC modifies the magnetization reversal behavior and nonvolatile electrical switching of QAH states.
To this end, we investigate QAH states and magnetism in tMBG devices proximitized by either a monolayer or bilayer of WSe\textsubscript{2}, which we refer to as tMBG/WSe\textsubscript{2}. 

We investigate five tMBG/WSe\textsubscript{2} devices (D1--D5) with twist angles in the range $\theta \approx 1.1 - 1.5\degree$ (Fig.~\ref{fig:S:DevicesOptical} and Tab.~\ref{tab:ListOfDevices}). In all these devices, WSe\textsubscript{2} is placed on the Bernal bilayer side (Fig.~\ref{fig:1}a).
Band structure calculations indicate that this placement of the WSe\textsubscript{2} layer induces stronger SOC at $D<0$ in the valence band and at $D>0$ in the conduction band, with the QAH and correlated metallic states observed in the latter regime (Fig.~\ref{fig:1}b).
The presence of proximity-induced SOC is confirmed by the measurements of quantum oscillations (Fig.~\ref{fig:S:SOCevidence} and Methods). 

Similar to tMBG devices without WSe\textsubscript{2}~\cite{polshyn_electrical_2020, chen_electrically_2021, he_competing_2021, xu_tunable_2021, polshyn_topological_2022, zhang_manipulation_2024, zhang_local_2023, waters_topological_2024, peng_abundant_2024}, tMBG/WSe\textsubscript{2} devices exhibit correlated metallic and gapped states at partial fillings of the conduction band for $D>0$ (Fig.~\ref{fig:S:DevicesResistance}). 
Measurements of longitudinal resistance ($R_{xx}$) and Hall resistance ($R_{yx}$) in devices D2, D3, and D4, with $1.2<\theta<1.4\degree$, reveal a large region of metallic isospin (spin and valley) ferromagnetic states surrounding an insulating state at moir\'e superlattice filling $\nu=2$ (Fig.~\ref{fig:1}c and Fig.~\ref{fig:S:DevicesResistance}). 
As highlighted by the $R_{yx}$ data~(Fig.~\ref{fig:S:DevicesResistance}), two smaller pockets of distinct metallic isospin-ferromagnetic states are centered near $\nu=1$ and $3$~\cite{polshyn_electrical_2020, chen_electrically_2021, he_competing_2021, polshyn_topological_2022, waters_topological_2024}. 
Devices D2 and D3, with twist angles $\theta=1.23\degree$ and $1.32\degree$, show QAH states at $\nu=1$ and $3$ (Fig.~\ref{fig:1} and Fig.~\ref{fig:S:D3_QAH}).
Despite the overall superficial similarity to non-proximitized tMBG, the studied tMBG/WSe\textsubscript{2} devices exhibit several distinct features: (i) well-quantized QAH states, (ii) an additional magnetization reversal in the metallic state $1<\nu<2$, and (iii) the absence of such a reversal near $\nu=3$. 

 \subsection{Quantized anomalous Hall effect}
In tMBG/WSe\textsubscript{2} device D2, QAH states at fillings $\nu=1$ and $3$ are visible as sharp dips in the longitudinal resistance $R_{xx}$ (Fig.~\ref{fig:1}c). 
A small magnetic field of 50~mT is applied during this measurement to suppress domain formation and random switching due to bistability. 
Measurements of the Hall resistance $R_{yx}$ as a function of temperature $T$ and displacement field $D$ reveal dome-shaped stability regions of the QAH states, marked by $R_{yx}\approx -0.5h/e^2$ (Fig.~\ref{fig:1}d,e).
The maximum onset temperatures are about 3.5~K and 6~K for $\nu=1$ and $3$, respectively, comparable to those reported in tMBG devices with similar twist angle but without WSe\textsubscript{2}~\cite{polshyn_electrical_2020}.
Both QAH states show magnetic hysteresis with $R_{yx}$ close to the quantized value (Fig.~\ref{fig:1}f,g).
For the QAH state at $\nu=3$, which has larger gap, as evidenced by its higher onset temperature, the zero-field $|R_{yx}|$ remains within $0.5\%$ of $0.5h/e^2$.
This robust quantization is reproduced across all three pairs of Hall contacts along the device channel (Fig.~\ref{fig:S:QAH}).
Moreover, the resistance measured in the configuration $R^*_{xx}$ (See Fig.~\ref{fig:1}f), which is topologically equivalent to the longitudinal resistance, is nearly zero. 
Such well-quantized QAH states were not previously observed in tMBG devices, likely because  of the residual twist-angle disorder~\cite{lau_reproducibility_2022}. 
In contrast, device D2 is highly homogeneous, with twist-angle variation $\delta \theta<0.02\degree$ across the device, as evidenced by two-terminal resistance measurements from different contact pairs (Fig.~\ref{fig:S:D2Homogeneity}).
This homogeneity is likely aided by the use of WSe\textsubscript{2} as a substrate~\cite{arora_superconductivity_2020} and by the improved fabrication procedure (Methods).

A quantized plateau is a defining hallmark of a quantum Hall state, arising from the interplay between topological edge transport and the disorder-induced localization.  
For the QAH state at $\nu=3$, we observe a robust quantized plateau as a function of both $n$ and $D$ (Fig.~\ref{fig:1}h and Fig.~\ref{fig:S:QAH}).
When measured as a function of magnetic field and carrier density, $R^*_{xx}$ shows a linear-in-field shift of the plateau position in $n$, consistent with the Streda formula behavior for a QAH with Chern number $C=\pm2$ (Fig.~\ref{fig:S:QAH}).
The derivative $dR^*_{xx}/dn$, plotted over a smaller range of $n$ and $B$ highlights the structure of this plateau (Fig.~\ref{fig:1}i). 
Outside the plateau, $R^*_{xx}$ exhibits slow variations, whereas inside the plateau, the signal is nearly featureless apart from small noise. 
This behavior confirms transport dominated by chiral edge states and negligible bulk contribution.

Typically, Hall plateau breakdown is associated with a disorder-driven localization-delocalization transition, in which a percolating conducting path connects opposite edges of the device~\cite{huckestein_scaling_1995}.
In our device, the plateau is flanked on both sides by intermediate regions containing sharp peaks closely spaced in $n$ (Fig.~\ref{fig:1}i).
These peaks disperse with magnetic field with the same slope as the QAH state, consistent with the Streda formula for a $C=2$ Chern insulator.
They disappear above $1$~K and are replaced by a finite resistance, at a temperature much lower than the QAH onset temperature (Fig.~\ref{fig:S:QAH}).
Given the micron-scale device dimensions, comparable to the length scale of twist-angle disorder, these features likely have a mesoscopic origin involving individual defects.
Indeed, the peak spacing corresponds to adding only a few electrons to the device region between the leads, suggesting that the peaks arise from single-electron charging of individual defects.
This interpretation is consistent with previous observations of defect charging in graphene moir\'e systems near superlattice insulating and field-induced Chern insulating states~\cite{tilak_flat_2021, dolleman_negative_2024, pierce_tunable_2025}.
Our findings show that moir\'e disorder defects can interfere with clean edge-state transport, posing an additional challenge for mesoscopic devices with quantum point contacts.

\subsection{Magnetization reversals}
We find that tMBG/WSe\textsubscript{2} devices exhibit a new magnetization reversal in the metallic regime between the QAH state at $\nu=1$ and the insulator at $\nu=2$ (Fig.~\ref{fig:2}a,b).
In device D2, several signatures indicate the presence of this reversal. 
First, the sense of the magnetic hysteresis in $R_{yx}$ changes near $\nu\approx1.25$, accompanied by a strong increase in the coercive field, indicating that the magnetization crosses zero and changes sign (Fig.~\ref{fig:2}c). 
Second, carrier-density sweeps show strong hysteresis, with two magnetic states characterized by distinct Hall resistances (Fig.~\ref{fig:2}d).
At higher fields, the $n$-hysteresis is narrow and centered near $\nu\approx1.25$, but it rapidly broadens as $B$ approaches zero, eventually spanning the range between $\nu\approx0.8$ and $1.6$.
This hysteretic response enables nonvolatile electrical switching of the $\nu=1$ QAH state (Fig.~\ref{fig:2}e).
Together, these observations establish the presence of a MR.

Alongside  magnetic-field and carrier-density hysteresis, which are universal features of MR switching, the MR is also accompanied by additional features that likely have a mesoscopic origin. 
As shown in Fig.~\ref{fig:2}a,b, sharp changes in both $R_{xx}$ and $R_{yx}$ appear over a narrow range of carrier density  near the MR. 
We attribute these features to an intermediate state in which a domain wall is trapped between the measurement leads, likely stabilized by residual disorder or device geometry. 
This interpretation is consistent with the strong contact dependence of these features (Fig.~\ref{fig:S:D3andD4switching}). 
Domain walls stabilized by higher magnetic  
fields are also likely to be responsible for the fine structure observed in Fig.~\ref{fig:2}c near $\nu=1.25$ and $|B|>0.5$~T.

\begin{figure}[t!]
\includegraphics[width=0.5\textwidth]{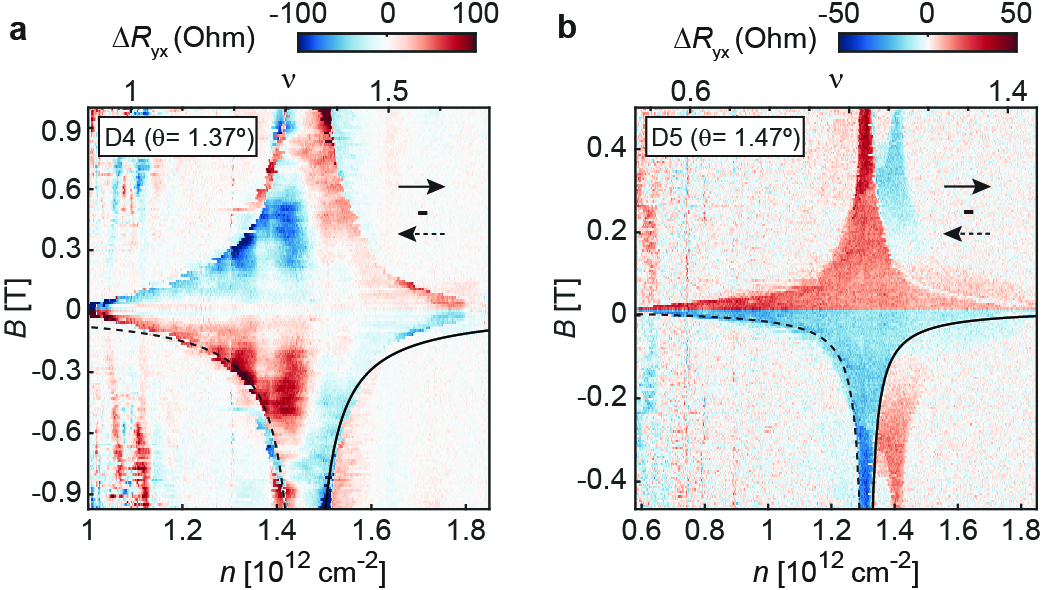} 
\caption{\textbf{Gate-driven magnetic state switching in tMBG/WSe\textsubscript{2} devices with higher twist angles.} 
\textbf{a, b,} 
The difference between $R_{yx}$ measured with two opposite $n$-sweep directions, $\Delta R_{yx}$, in tMBG/WSe\textsubscript{2} devices D4 ($\theta= 1.37\degree$) and D5 ($\theta= 1.47\degree$).  
Here, $R_{yx}$ is antisymmetrized with respect to $B$.
Device D4 is measured at $D=0.57$~V/nm, whereas in D5, $D$ spans $0.48-0.63$~V/nm, as indicated in Fig.~\ref{fig:S:D3andD4switching}.
Solid and dashed hyperbolic lines near the MR transitions are guides to the eye following $ B \sim 1/(n - n_{rev})$, where $n_{rev}$ is the carrier density at which MR occurs. 
 }
\label{fig:3}
\end{figure}

We find similar MR-driven switching at $1<\nu<2$ in all four studied tMBG/WSe\textsubscript{2} devices with twist angles $1.2\degree<\theta<1.5\degree$ (D2--D5).
The reversals are evidenced by features in the $(n,D)$ resistivity maps (Fig.~\ref{fig:S:DevicesResistance}) and by hysteresis of the magnetic states in carrier-density sweeps (Fig.~\ref{fig:3}a,b and Fig.~\ref{fig:S:D3andD4switching}). 
The MRs occur even when the gapped states at integer superlattice fillings in the conduction band are weak or absent. 
Device D4 shows only a weakly developed resistive state at $\nu=2$, while such states are absent in D5.
These observations indicate that the mechanism behind these MRs is robust and is only weakly dependent on the details of the band structure.

\begin{figure}[h!]
\includegraphics[width=0.5\textwidth]{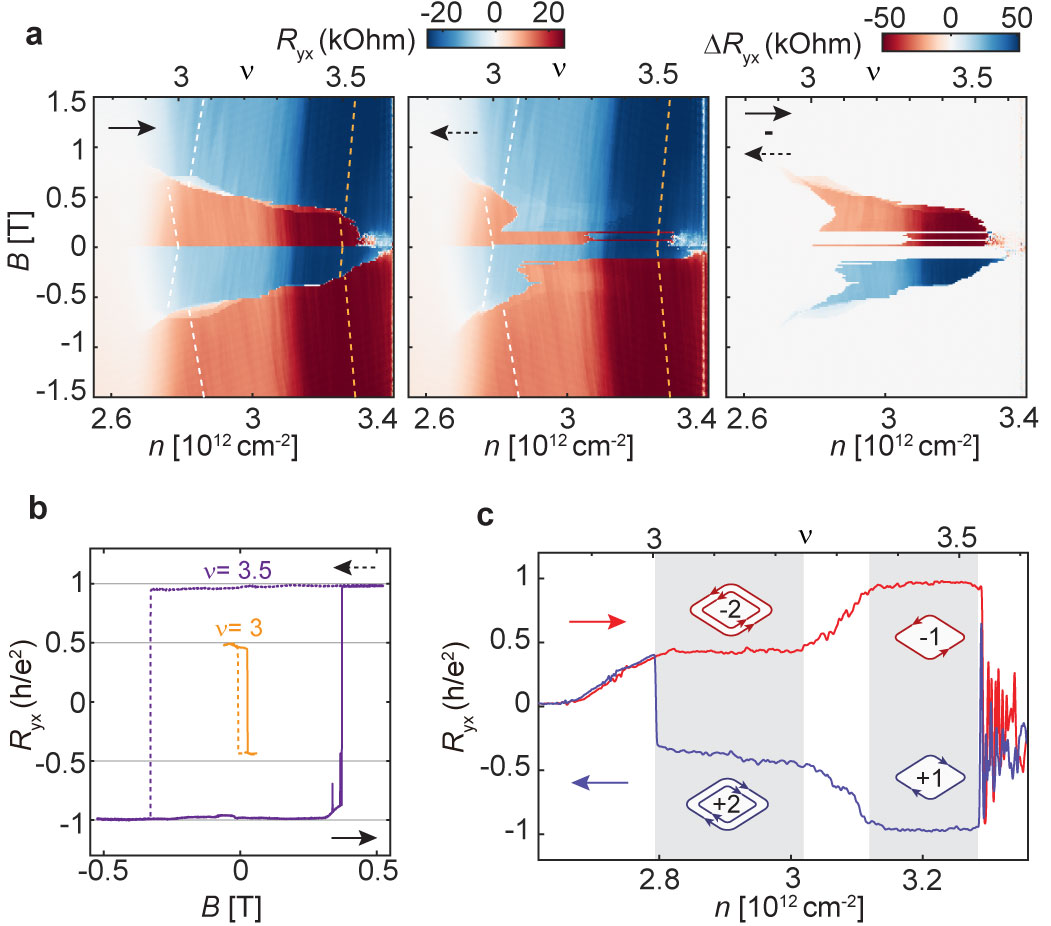} 
\caption{\textbf{Gate-tunable Chern number in electrically switchable QAH states in tMBG.} 
\textbf{a,} $R_{yx}$ measured as a function of $n$ and $B$ in device D6, which does not contain WSe\textsubscript{2}($1.27\degree$), at $D$ = 0.46~V/nm. 
The left and middle subpanels show the resistance measured for different $n$-sweep directions, while the right one plots the difference, $\Delta R_{yx}$. 
Dashed lines originating from $\nu=3$ and 3.5 have slopes expected for Chern insulators with $C = \pm 2$ and $C= \pm 1$, respectively. 
\textbf{b,} QAH states at $\nu$ = 3 and $\nu$ = 3.5 in D6 evidenced by the quantized hysteresis of $R_{yx}$.  
\textbf{c,} Nonvolatile gate-driven switching of the magnetic state at $B$ = 25~mT. 
Each magnetic state can be additionally gate-tuned between two plateaus corresponding to $|C|=2$ and $1$ Chern insulators. 
}
\label{fig:4}
\end{figure}

There is a strong contrast between the MRs observed in tMBG/WSe\textsubscript{2} and pristine tMBG devices with twist angles $\theta=1.2-1.3\degree$. 
Pristine tMBG devices show robust electrical switching due to an MR \textit{within} the $\nu=3$ QAH plateau~\cite{polshyn_electrical_2020, zhang_local_2023, zhang_manipulation_2024}, with the magnetic state switching from $C=-2$ to $C=+2$ as $n$ crosses the $\nu=3$ plateau at $B>0$.
By contrast, in tMBG/WSe\textsubscript{2} devices, the $C=+2$ magnetic state remains stable throughout the vicinity of $\nu=3$, with no signs of an MR (Fig.~\ref{fig:S:D2_switching}).
Conversely, pristine tMBG devices show no magnetization reversals in the metallic regime $1<\nu<2$ ~\cite{polshyn_electrical_2020, zhang_local_2023, polshyn_topological_2022}, where we observe them in tMBG/WSe\textsubscript{2}.
We further characterize a pristine tMBG device, D6, with $\theta=1.27\degree$, which confirms switching near $\nu=3$ but not in the metallic regime near $\nu=1$ (Fig.~\ref{fig:4}c,d).
We therefore conclude that WSe\textsubscript{2} proximity removes the MR near $\nu=3$ and induces a new MR at $1<\nu<2$, demonstrating its strong effect on the magnetic properties of tMBG.

\subsection{The effect of proximity-induced Ising SOC on the MRs}
Proximity-induced SOC is likely the dominant mechanism by which WSe\textsubscript{2} modifies the magnetic properties and gate-induced switching in tMBG. 
Although WSe\textsubscript{2} may also produce small changes in orbital magnetization through substrate-induced modifications of the electronic structure, its main effect is to enforce spin-valley alignment through SOC and thereby change the energy landscape of competing states.
In pristine tMBG, the intrinsic SOC is expected to be weak and even a small magnetic field fixes the spin polarization. As such, the competing magnetic states involved in gate-driven switching have the same spin polarization, and the switching is effectively confined to the valley sector, with the reversal in the orbital part of the magnetization controlling the switching condition. 
By contrast, proximity-induced SOC lifts the spin degeneracy of a valley-polarized magnetic state, making spin magnetization relevant to the switching process. As orbital and spin magnetizations have similar magnitudes in tMBG, they combine in non-trivial ways, thereby modifying the MRs.

To illustrate this mechanism, we use a simple model: SOC is neglected in pristine tMBG, while tMBG/WSe\textsubscript{2} includes Ising SOC, $\lambda_I \tau_z s_z/2$, with $\tau_z$ and $s_z$ Pauli matrices for valley and spin. 
At $\nu=3$, one moir\'e subband remains unoccupied, and valley polarization into $K$ or $K'$ gives QAH states with $C=+2$ or $-2$. 
In pristine tMBG at moderate fields $B>0$, switching occurs between states with either the $K\downarrow$ or $K'\downarrow$ flavor unoccupied, so the reversal is purely valley-like, and magnetic reversal occurs when the magnetization $M^K_\text{orbital}$ changes sign. 
In tMBG/WSe\textsubscript{2}, however, $\lambda_I>0$ favors states with either $K\uparrow$ or $K'\downarrow$ unoccupied, so the switching reverses both valley and spin, and magnetic reversal occurs when $M_{\mathrm{orbital}}^K+M_{\mathrm{spin}}$ changes sign. 
The possible scenario is illustrated in Fig.~\ref{fig:2}f and g, where $M_{\mathrm{orbital}}^K$ changes sign across $\nu = 3$ without $M_{\mathrm{orbital}}^K+M_{\mathrm{spin}}$ changing sign, consistent with the observation of MR at $\nu = 3$ only without the proximity of WSe\textsubscript{2}. 
Because the relative sign of $M_{\mathrm{spin}}$ and $M_{\mathrm{orbital}}$ depends on $\lambda_I$, Ising SOC can either favor or suppress an MR. 
Applying the same argument at $\nu=1$ reverses the sign of the spin contribution. 
As illustrated in Fig.~\ref{fig:2}g, this tends to favor an MR, though it depends on the magnitude of $M_{\mathrm{orbital}}^K$. 

The MR observed in the metallic regime $1<\nu<2$ is more subtle as the evolution of orbital magnetization for metallic states, as compared to that across a Chern gap~\cite{zhu_voltage_2020}, follow less universal patterns.
Nevertheless, the MR observed in the metallic regime $1<\nu<2$ can trace its origin back to the SOC-enforced spin-valley alignment, which reduces the net magnetization and make reversal points easier to stabilize. We perform microscopic magnetization calculations to support this scenario (SI).

The magnitude and sign of proximity-induced SOC can be tuned by the rotational alignment between graphene and the TMD~\cite{zhang_twist-programmable_2025}. 
Thus, proximity-induced SOC provides a route to engineering magnetization through the relative alignment of orbital and spin magnetic moments. 
This approach may enable control of the sign and strength of the coupling to magnetic field for a broad class of graphene moir\'e systems, including integer and fractional QAH states in rhombohedral graphene multilayers~\cite{lu_fractional_2024}.

\subsection{Tuning metastability and Chern number of magnetic states}
Finally, we discuss the metastability of magnetic states in tMBG and the functionality it can enable.
The hysteresis of magnetic states in $n$-sweeps shows a strong magnetic-field dependence in all devices.
We find that the metastable range in $n$ closely follows a ${\sim|B|^{-1}}$ dependence, as shown by the guide lines in Fig.~\ref{fig:2}f and Fig.~\ref{fig:3}a,b.
This behavior is consistent with a simple picture in which metastability is lost when the magnetic-energy change, $B\cdot M(n)$, exceeds a fixed energy barrier, assuming that $M(n)$ varies approximately linearly with $n$ near the MR point.
Thus, the metastable range can be tuned over a wide carrier-density interval, reaching nearly one electron per moir\'e unit cell at low fields.

Strong low-field metastability allows magnetic states to be tuned across distinct regimes without resetting. 
In device D2 at $B=-23$~mT, for example, the system can be tuned between QAH states with $C=\pm2$ at $\nu=1$ and a metallic magnetic state near $\nu\approx1.3$ with vanishing Hall resistance (Fig.~\ref{fig:2}e). 
For potential QAH memory applications, this tunability could enable addressing an individual QAH memory bit in a series-connected chain of bits.

Strong metastability also allows tuning between QAH states with distinct Chern numbers. 
In tMBG device D6, besides the QAH state at $\nu=3$ with $C=\pm2$, we find a well-quantized QAH state at $\nu=3.5$ with $C=\pm1$ (Fig.~\ref{fig:4}a,b), consistent with a previously observed non-quantized version of this state~\cite{polshyn_topological_2022}. 
The appearance of a quantized response at non-integer moir\'e band filling is attributed to  moir\'e translational symmetry-breaking. 
Because both states emerge from the same magnetic pocket, sweeping the carrier density from $\nu\sim3$ to $\nu\sim3.5$ tunes the system between $|C|=2$ and $|C|=1$ without resetting the magnetic state (Fig.~\ref{fig:4}c). 
Thus, the MR enables nonvolatile switching of the magnetic state, while metastability allows this state to be tuned between different plateaus.
Together, these capabilities demonstrate the use of a single control parameter to tune the number of conducting chiral edge channels, alongside nonvolatile control of edge-state chirality, which could enable new architectures for reconfigurable chiral-edge-state devices~\cite{ovchinnikov_topological_2022}.

\clearpage


%

\section*{Methods}
\subsection*{Device fabrication}
 tMBG heterostructures were assembled using a dry transfer method with a polydimethylsiloxane (PDMS) stamp covered by a thin polycarbonate (PC) film. 
Exfoliated thin crystals were picked up in the following sequence, from top to bottom: FLG-hBN-WSe\textsubscript{2}-BLG-MLG-hBN-FLG, where FLG is few-layer graphene, hBN is hexagonal boron nitride, BLG is Bernal-stacked bilayer graphene, and MLG is monolayer graphene. 
In our tMBG device, the WSe\textsubscript{2} and BLG were located in the top part of the heterostructure, further away from the Si substrate. 
However, in the main text, we use the opposite convention, as shown in Fig.~\ref{fig:1}a, to remain consistent with earlier tMBG studies and the definition of $D$.
To reduce disorder and prevent twist-angle relaxation, we used thin top hBN ($<20$~nm) and aligned the edges of the graphene and top hBN~\cite{diez-merida_high-yield_2025}. 
During stacking, most crystals were picked up at approximately $100\degree$~C, except for the MLG and BLG layers, which were picked up at $40\degree$~C.
Completed heterostructures were released at 180\textdegree C onto a SiO\textsubscript{2}/Si substrate.  The heterostructures were then processed using electron-beam lithography, CHF\textsubscript{3}/O\textsubscript{2} etching, and metal deposition to form edge contacts with Cr/Pd/Au thicknesses of 3/15/150 nm, respectively.

\subsection*{Device characterization}
 Most measurements were performed in a cryogen-free dilution refrigerator (LD250 Bluefors) with the nominal mixing chamber base temperature of about 20~mK. 
 The measurement lines have low-temperature filters with frequency cut-ff frequency of 65~kHz.
 Transport measurements at higher temperature, in particular those in Fig.~\ref{fig:1}d,e and Fig.~\ref{fig:S:QAH}k, were performed in a cryogen-free system (TesltronPT Oxford Instruments) with a variable temperature insert or a helium-3 insert, which is also equipped with low-temperature filters. 
 
 The resistance was measured using voltage preamplifiers (SRS), a current preamplifier (DL Instruments), and lock-in amplifiers (SRS) with excitation currents of 0.5-5~nA at frequency of 17.77~Hz. 
 The gate voltages were applied using a custom DAC-ADC (openDACs).
 The carrier density, $n = (c_t \mathrm{v}_t + c_b \mathrm{v}_b)/e$, and electric displacement field, $D = (c_t \mathrm{v}_t- c_b \mathrm{v}_b)/2\varepsilon_0$, were tuned independently by controlling the top - and bottom-gate voltages, $ \mathrm{v}_t$  and $\mathrm{v}_b$, respectively, where $c_t$ ($c_b$) is the capacitance per unit area of the top (bottom) gate, $e$ is the elementary charge, and $\varepsilon_0$ is the vacuum permittivity.  
 $c_t$  and $c_b$ are determined by fitting the Landau fan features of the devices.  
The twist angles in devices D1, D4, and D5 were determined from the  carrier densities at which of the superlattice insulating states ($\nu=\pm4$) were observed: $n_s = 8\sin^2 \theta/{\sqrt{3}}a^2$, where $a = 0.246$~nm is the graphene lattice constant.
For devices D2, D3 and D6, we determined the twist angle from the  position of the $\nu=2$ correlated state observed at $0.5n_s$.

\subsection*{Proximity-induced spin-orbit coupling}
To confirm the presence of SOC in tMBG/WSe\textsubscript{2}, we perform quantum oscillation measurements in the valence band, which is more dispersive and does not show apparent signs of correlated phases (Fig.~\ref{fig:S:SOCevidence}).
We find splitting of the Fermi surfaces for $D<0$ but not for $D>0$, which is consistent with the band structure calculations, shown in Fig.~\ref{fig:1}b.
Using the density of states obtained from electronic structure calculations, we estimate the corresponding energy splitting to be $\sim1$~meV, which is consistent with the previous studies of   graphene systems proximitized with WSe\textsubscript{2} \cite{island_spinorbit-driven_2019, zhang_twist-programmable_2025}, where it was attributed to the Ising SOC $H_I=0.5\lambda_I \tau_z s_z$, with $\tau_z$ and $s_z$ denoting the Pauli matrices for the valley and spin degrees of freedom, respectively.   
While proximity-induced Rashba SOC is also possible, calculations show that it has a relatively weak effect on the band structure compared with Ising SOC (see Supplementary Information).
Most importantly for our interpretation of the influence of SOC on MR-driven switching, even when both Ising and substantial Rashba spin-orbit coupling are present, the conduction band at $D > 0$ still develops substantial out-of-plane spin polarization that is locked to the valley.

\subsection*{Calculations of the band structure and magnetization}
We adopt a continuum description for tMBG, which is a generalization of the Bistritzer-MacDonald model~\cite{Bistritzer2011}. Substrate-induced SOC is taken to be non-zero only on the graphene layer immediately adjacent to the WSe\textsubscript{2}. We perform self-consistent Hartree-Fock to capture the effect of electron-electron interaction, and orbital magnetization is then calculated based on the mean-field bandstructure obtained from Hartree-Fock calculations. Details of the model are discussed in the SI.

\section*{acknowledgments}
The authors are grateful to Jihang Zhu for fruitful discussions and sharing calculations that motivated this project. 
We also acknowledge discussions with  Fangyuan Yang,  Thomas Scaffidi, and Allan MacDonald. 
Work done at ISTA is supported by ERC grant (OCI, 101118064). Views and opinions expressed are however those of the authors only and do not necessarily reflect those of the European Union or the European Research Council Executive Agency. Neither the European Union nor the granting authority can be held responsible for them.
 Z.W. acknowledges support from a Leverhulme Trust International Professorship (Grant Number LIP-202-014). 
 K.W. and T.T. acknowledge support from the CREST (JPMJCR24A5), JST and World Premier International Research Center Initiative (WPI), MEXT, Japan.
 S.H.S. acknowledges support from the UK Engineering and Physical Sciences Research Council (Grant No. EP/X030881/1). S.A.P. acknowledges support from a  UKRI  Frontier Research Grant (Grant No. EP/Z002419/1, Guarantee for an ERC Consolidator Grant). 
 This project was also supported by the Scientific Service Units of ISTA through resources provided by the Nanofabrication facility and the
MIBA Machine Shop.

\section*{Author contributions}
M.Q. fabricated the devices and performed the measurements, advised by H.P.  
K.W. and T.T. grew the hBN crystals. 
G.K. built the low-temperature filters used in the transport measurements. 
Z.W., S.H.S., and S.A.P. contributed to the theoretical interpretation and performed band structure calculations. 
H.P., M.Q. and Z.W. wrote the manuscript with inputs from all other authors. 

\section*{Competing interests}
The authors declare no competing interests.

\section*{Data availability}
The data that support the plots within this paper and other findings of this study are available from the corresponding author upon reasonable request.

\renewcommand{\figurename}{\textbf{Extended Data Fig.}}
\renewcommand{\thefigure}{E\arabic{figure}}
\setcounter{figure}{0}
\renewcommand{\tablename}{\textbf{Extended Data Tab.}}
\renewcommand{\thetable}{E\arabic{table}}
\setcounter{table}{0}

\newpage\clearpage

\widetext

\begin{table}[ht]
    \centering
    \begin{tabular}{|c|c|c|}
    \hline
         Device & $\theta$ & WSe\textsubscript{2}  \\[15pt]
    \hline \hline
        D1 (MQ34)& $1.09\degree$ & monolayer/BLG side  \\[15pt]   
    \hline
        D2 (MQ30)& $1.23\degree$ & monolayer/BLG side  \\[15pt]
     \hline 
        D3 (MQ25)& $1.32\degree$ & monolayer/BLG side  \\[15pt]
    \hline
        D4 (MQ39)& $1.37\degree$  & bilayer/BLG side   \\[15pt]
    \hline
        D5 (MQ36)& $1.47\degree$ & bilayer/BLG side    \\[15pt]
    \hline
        D6 (MQ46)& $1.27\degree$  & none \\[15pt]
     \hline 
    \end{tabular}
    \caption{tMBG devices investigated in this study.}
    \label{tab:ListOfDevices}
\end{table}

\begin{figure*}[ht!]
\includegraphics[width=0.8\textwidth]{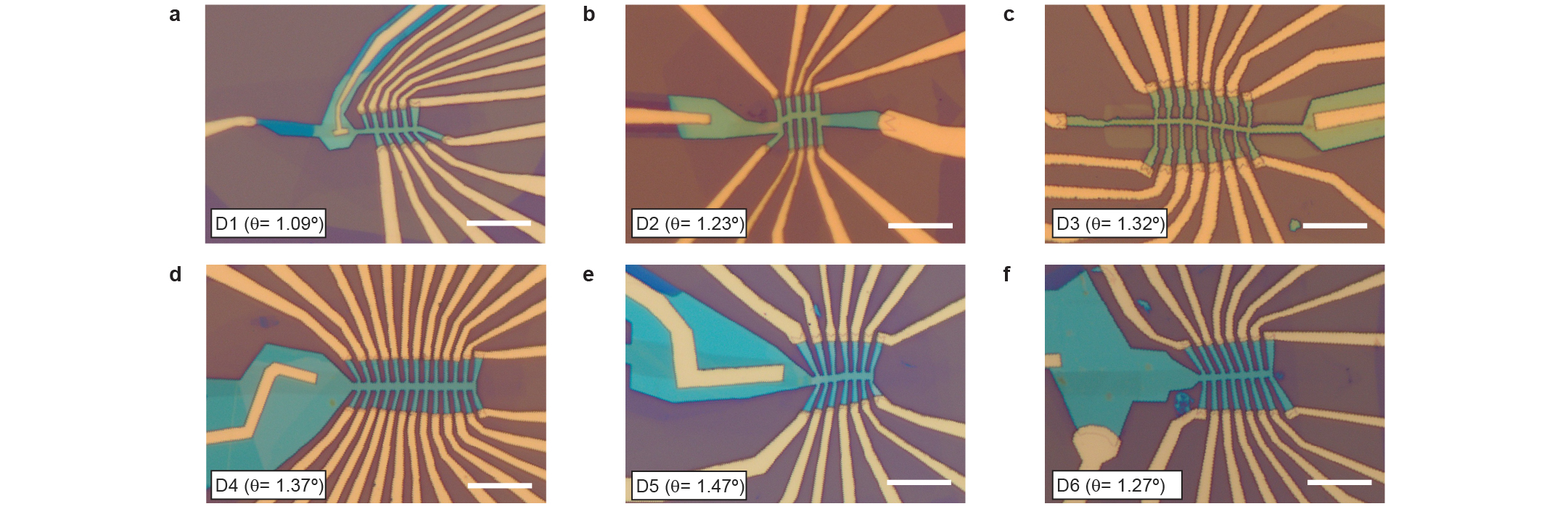} 
\caption{\textbf{Optical images of the tMBG devices.} Scale bars are 10~$\mathrm{\mu m}$. }
\label{fig:S:DevicesOptical}
\end{figure*}

\begin{figure}[ht!]
    \centering
    \includegraphics[width=\textwidth]{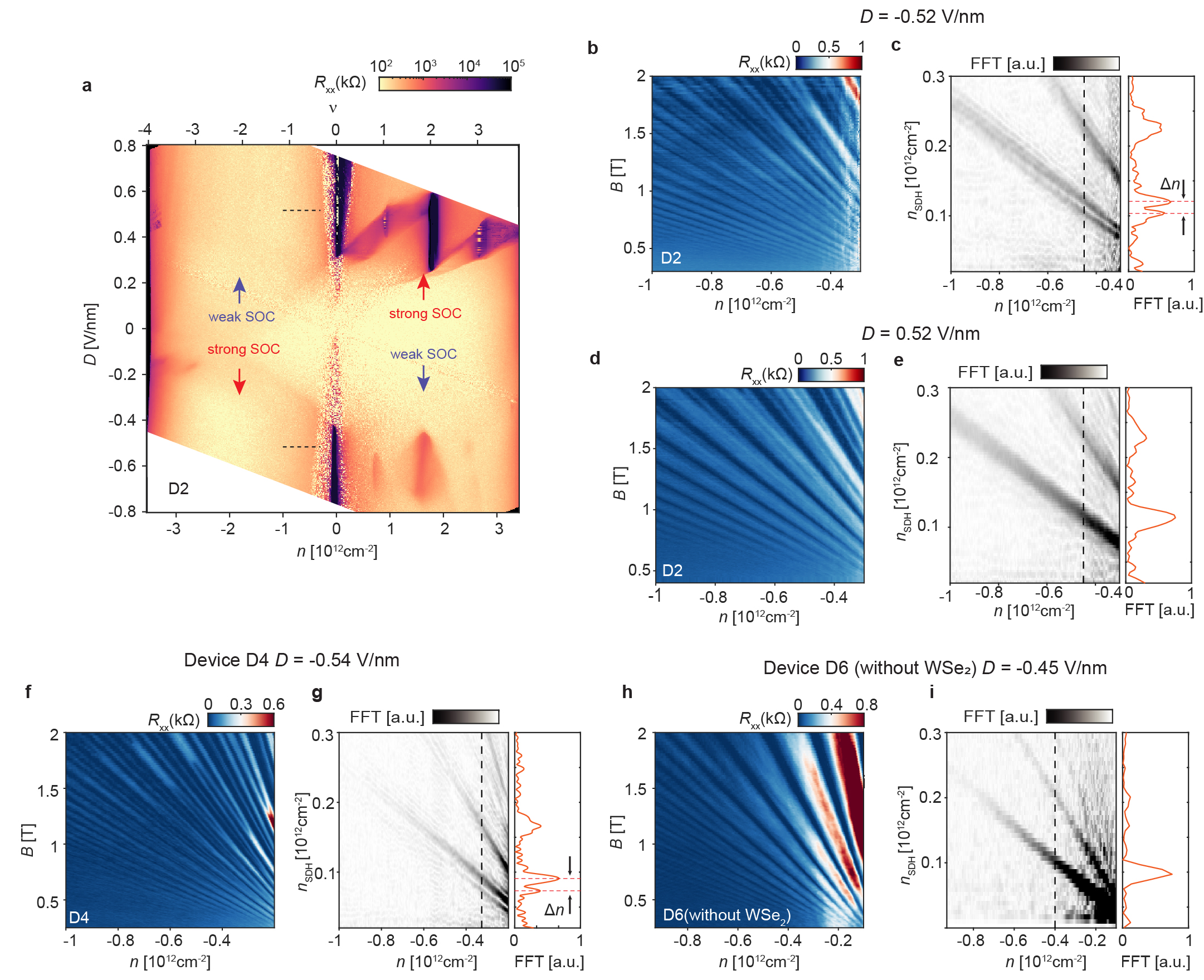}
    \caption{
    \textbf{Evidence for proximity-induced SOC.} \textbf{a}, Full-range map of $R_{xx}$ in device D2 as a function of $D$ and $n$ at zero magnetic field. 
    Because WSe\textsubscript{2} is placed on the bilayer-graphene side of the device, stronger SOC is expected in the valence band at $D<0$ and in the conductance band at $D>0$, as indicated. 
    \textbf{b},\textbf{d}, Quantum oscillations in $R_{xx}$, measured in the valence band at $D$ = -0.52 V/nm (\textbf{b}) and $D$ = 0.52 V/nm (\textbf{d}). The data were acquired along the trajectories marked by dashed lines in \textbf{a}. 
    \textbf{c},\textbf{e}, Fast Fourier transforms (FFTs) of $R_{xx}$ in \textbf{b} and \textbf{d}, with frequencies converted to equivalent carrier densities, $n_{SDH}$. 
    Side panels show line cuts though the FFT maps along the dashed lines in the main panels. 
    At $D>0$, where stronger SOC is expected in the valence band, a splitting between the Fermi-surface densities, $\Delta n$, is observed.
    Using the density of states obtained from band-structure calculations, we estimate the corresponding energy splitting to be $\sim0.7$~meV, consistent with the typical magnitude of proximity-induced Ising SOC  reported in other graphene systems.  
    \textbf{f, g}, Analogous analysis for device D4. Quantum oscillations were measured at $D$ = -0.54 V/nm. The observed splitting between the Fermi-surface densities provide evidence for SOC. 
    We estimate the corresponding energy splitting that corresponds to be $\sim0.9$meV.
     \textbf{h, i}, Quantum oscillations measured in a comparable hole-doped regime ($D$ = -0.45 V/nm) in D6, which does not contain WSe\textsubscript{2}, show no resolvable splitting due to SOC. 
    }
    \label{fig:S:SOCevidence}
\end{figure}

\begin{figure*}[ht!]
\includegraphics[width=\textwidth]{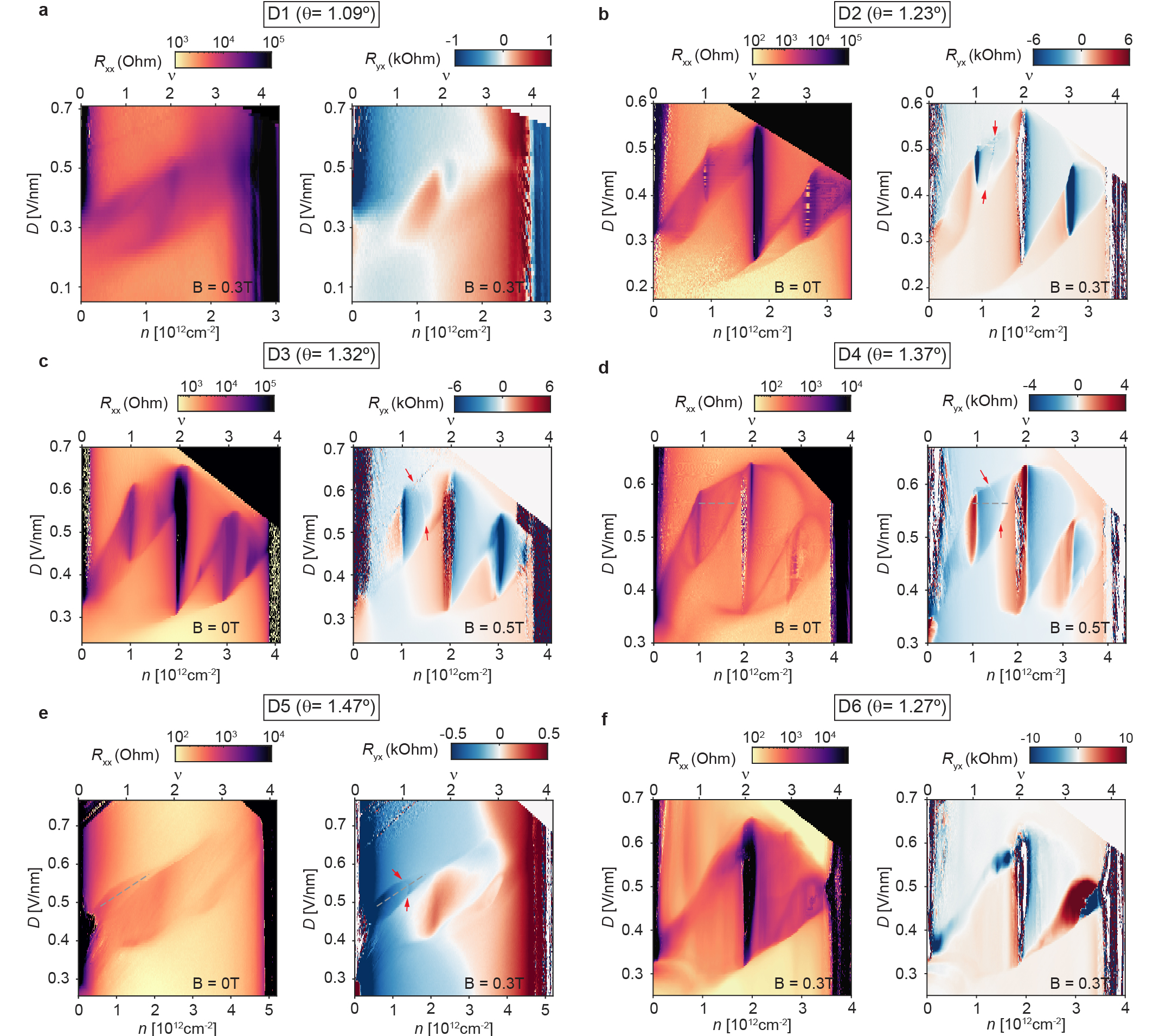} 
\caption{\textbf{Longitudinal ($R_{xx}$) and Hall ($R_{yx}$) resistances of tMBG devices.} 
\textbf{a-f}, $R_{xx}$ and $R_{yx}$ measured in the region exhibiting correlated states ($D>0$, $n>0$) in all devices studied. 
The magnetic fields used for the measurements are indicated in the respective panels.
All maps measured at nonzero magnetic field were symmetrized ($R_{xx}$) or antisymmetrized ($R_{yx}$) with respect to the magnetic field. 
Red arrows in $R_{yx}$ maps in panels \textbf{b--e} mark gate-driven magnetic state switching events caused by magnetization reversals. 
All data were acquired at 20~mK, except for the $R_{xx}$ and $R_{yx}$ data in \textbf{a} and the $R_{xx}$ data in \textbf{d}, which were measured at 1.5~K.} 
\label{fig:S:DevicesResistance}
\end{figure*}

\begin{figure*}[ht!]
\includegraphics[width=0.4\textwidth]{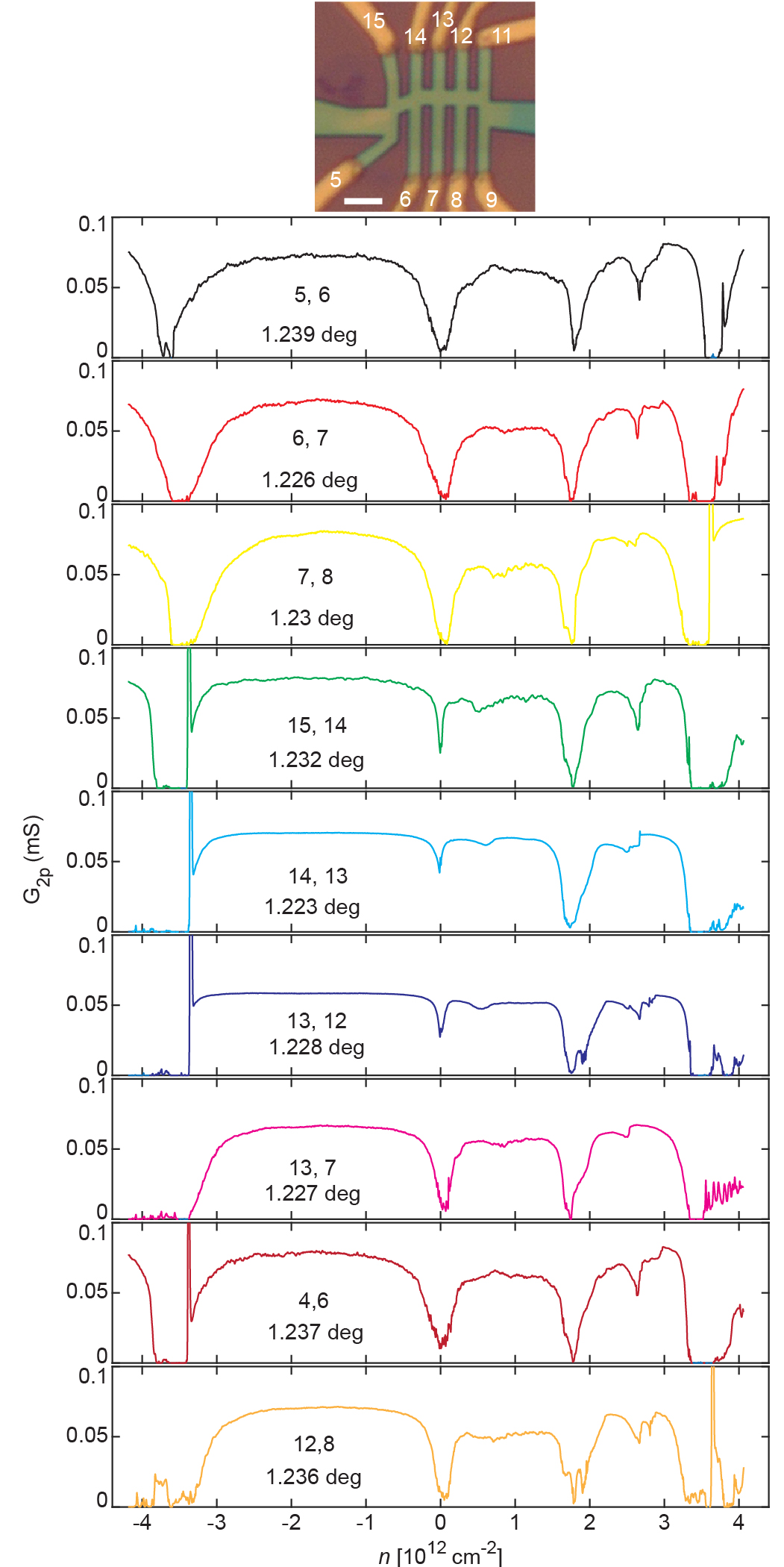} 
\caption{\textbf{Twist angle disorder in D2.} 
Two-terminal conductance measured using different contact-pair combinations, as labeled in the panels. 
The twist angles in different sections of the device, estimated from the positions of the $\nu = 2$ conductance minima, indicate a twist-angle variation of less than $0.02 \degree$. }
\label{fig:S:D2Homogeneity}
\end{figure*}

\begin{figure*}[ht!]
\includegraphics[width=\textwidth]{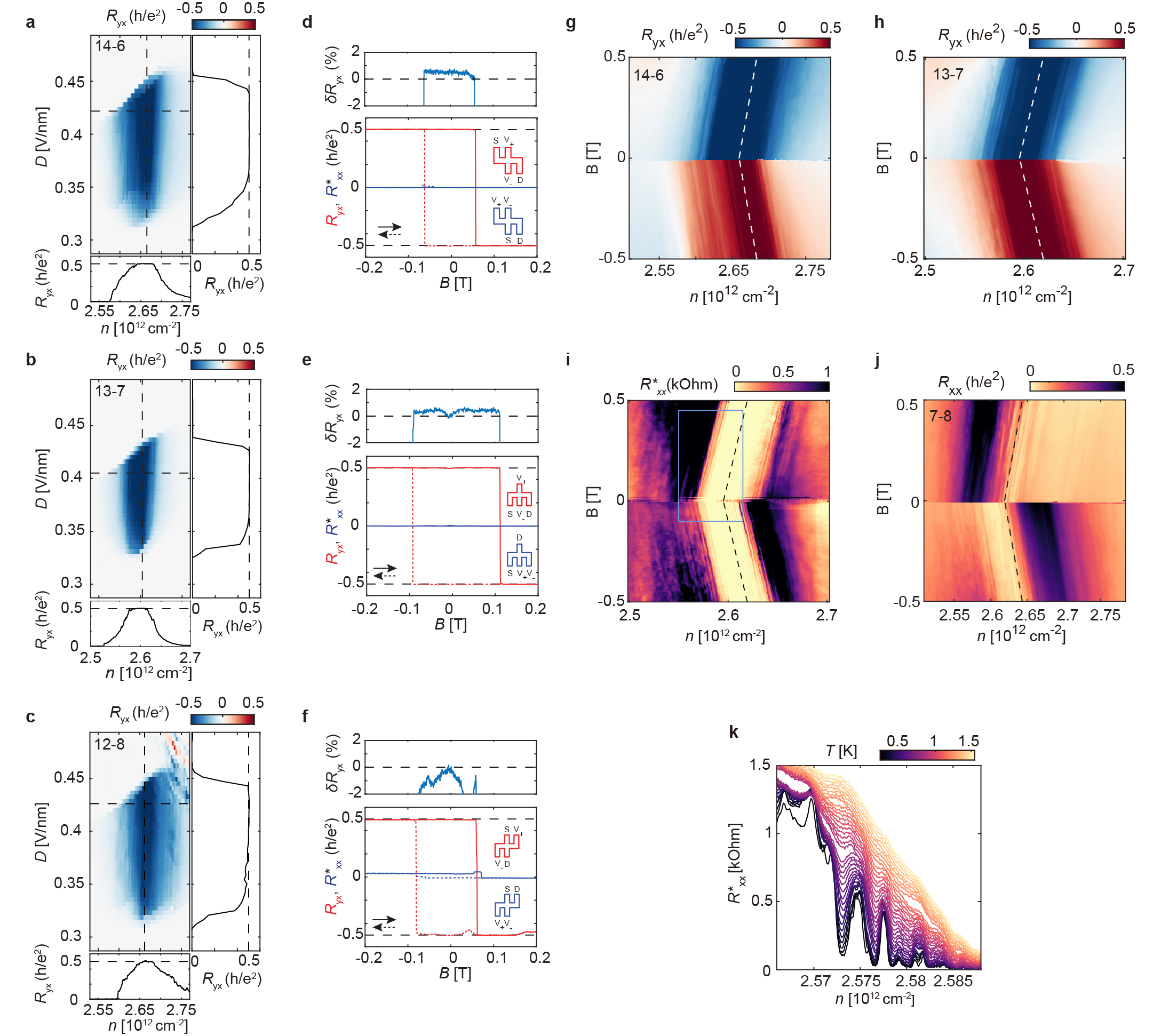} 
\caption{\textbf{Quantized anomalous Hall effect at $\nu$= 3 across device D2.} 
\textbf{a,b,c}, Antisymmetrized ${R}_{yx}$ as a function of \textit{D} and \textit{n} at $B$= 100~mT measured using the indicated contact pairs. The line cuts of ${R}_{yx}$ shown along the bottom and right sides of the main plots are taken at the positions inidcated by dashed lines.  
\textbf{d,e,f}, Magnetic hysteresis of ${R}_{yx}$ and $R^*_{xx}$. The measurements were performed at the points that correspond to the intersections of dashed lines in panels \textbf{a,b,c}. Top panels show the deviation from the expected quantized value,  $\delta R_{yx} = h^{-1} e^{2}({R}_{yx} (B_{up}) - {R}_{yx} (B_{down})-h e^{-2})$. 
\textbf{g,h}, $R_{yx}$ as a function of $n$ and $B$ measured between the indicated contact pairs. 
\textbf{i}, $R^*_{xx}$ measured in the same contact configuration as in (e), as a function of $n$ and $B$. The rectangle indicates the range shown in Fig.~\ref{fig:1}\textbf{i}. 
\textbf{j}, $R_{xx}$ as a function of \textit{n} and \textit{B} measured between the indicated contact pair. 
The dashed lines \textbf{g--j} calculated using the Streda formula correspond to Chern numbers $C = \pm2$. 
\textbf{k}, Temperature dependence of $R^*_{xx}$  measured at the edge of the QAH plateau.}
\label{fig:S:QAH}
\end{figure*}

\begin{figure*}[ht!]
    \centering
    \includegraphics[width=\textwidth]{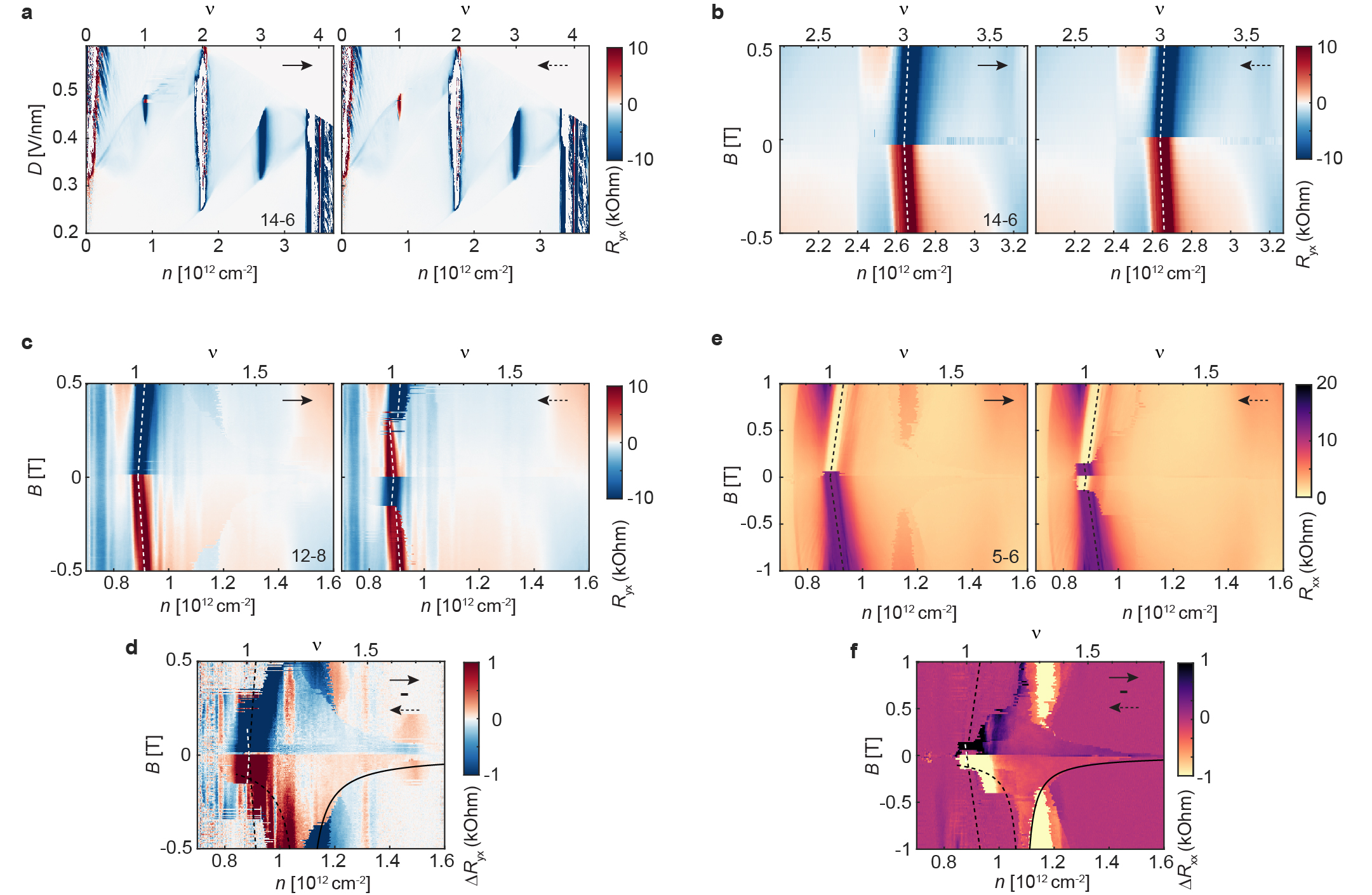}
    \caption{
    \textbf{Gate-driven switching of the QAH state at $\nu$= 1 and the absence of switching at $\nu$= 3 in D2.} 
    \textbf{a}, \(\textit{R}_{yx}\) as a function of  \textit{D} and \textit{n} at $B$= 50~mT as \textit{n} measured while sweeping $n$ in  the forward (left) and backward (right) directions. The QAH state at $\nu=1$ exhibits switching, whereas the state at $\nu=3$ does not.
    \textbf{b}, $R_{yx}$ measured near $\nu$ = 3 at $D$ = =0.39~V/nm while sweeping $n$ in opposite directions. No hysteresis or switching is observed. 
    \textbf{c,d}, Measurements of $R_{yx}$ analogous to those  shown in Fig.~\ref{fig:2}d but performed using a different pair of Hall-voltage probes (contacts 12 and 8), also reveals switching of the magnetic state and QAH states at $\nu=1$. $R_{yx}$ was measured and $B$ at $D$ = 0.47~V/nm. 
    \textbf{e,f}, Gate-driven switching near $\nu=1$ as observed in $R_{xx}$. The measurements performed at $D$ = 0.47~V/nm, using contacts 5 and  6 as voltage probes. 
    }
    \label{fig:S:D2_switching}
\end{figure*}

\begin{figure*}[ht!]
    \centering
    \includegraphics[width=\textwidth]{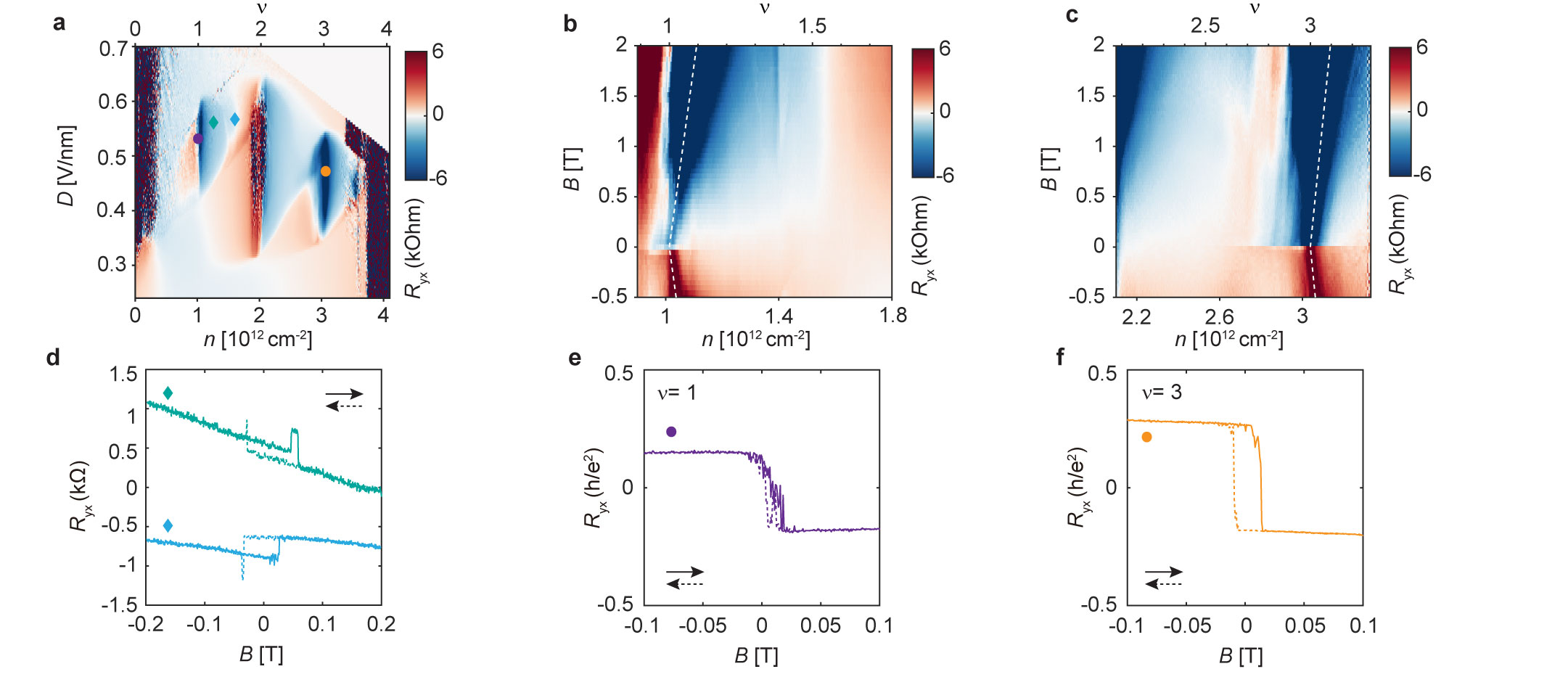}
    \caption{
    \textbf{Additional data on QAH states and magnetization reversal near $\nu$= 1 in device D3.} 
    \textbf{a}, Antisymmetrized $R_{yx}$ as a function of $D$ and $n$ at $B$= 300~mT. 
    \textbf{b, c}, Magnetic-field dependence of $R_{yx}$ measured near $\nu=1$ and $\nu=3$, respectively. In \textbf{b}, $D$ ranges from 0$.51$~V/nm to $0.64$~V/nm, whereas \textbf{c} was measured at $D=0.45$~V/nm.
    The QAH states at $\nu=1$ and $3$  are evidenced by a finite Hall resistance that persists down to zero magnetic field. 
    The dashed lines, calculated from the Streda formula, correspond  to Chern number $C = \pm2$. 
     \textbf{d--f}, Magnetic hysteresis of $R_{yx}$ measured across the magnetization reversal (d) and at $\nu=1$ (e), and $\nu=3$ (f). 
     We attribute the significant deviations from the quantized values to disorder.
    }
    \label{fig:S:D3_QAH}
\end{figure*}

\begin{figure}[ht!]
    \centering
    \includegraphics[width=\textwidth]{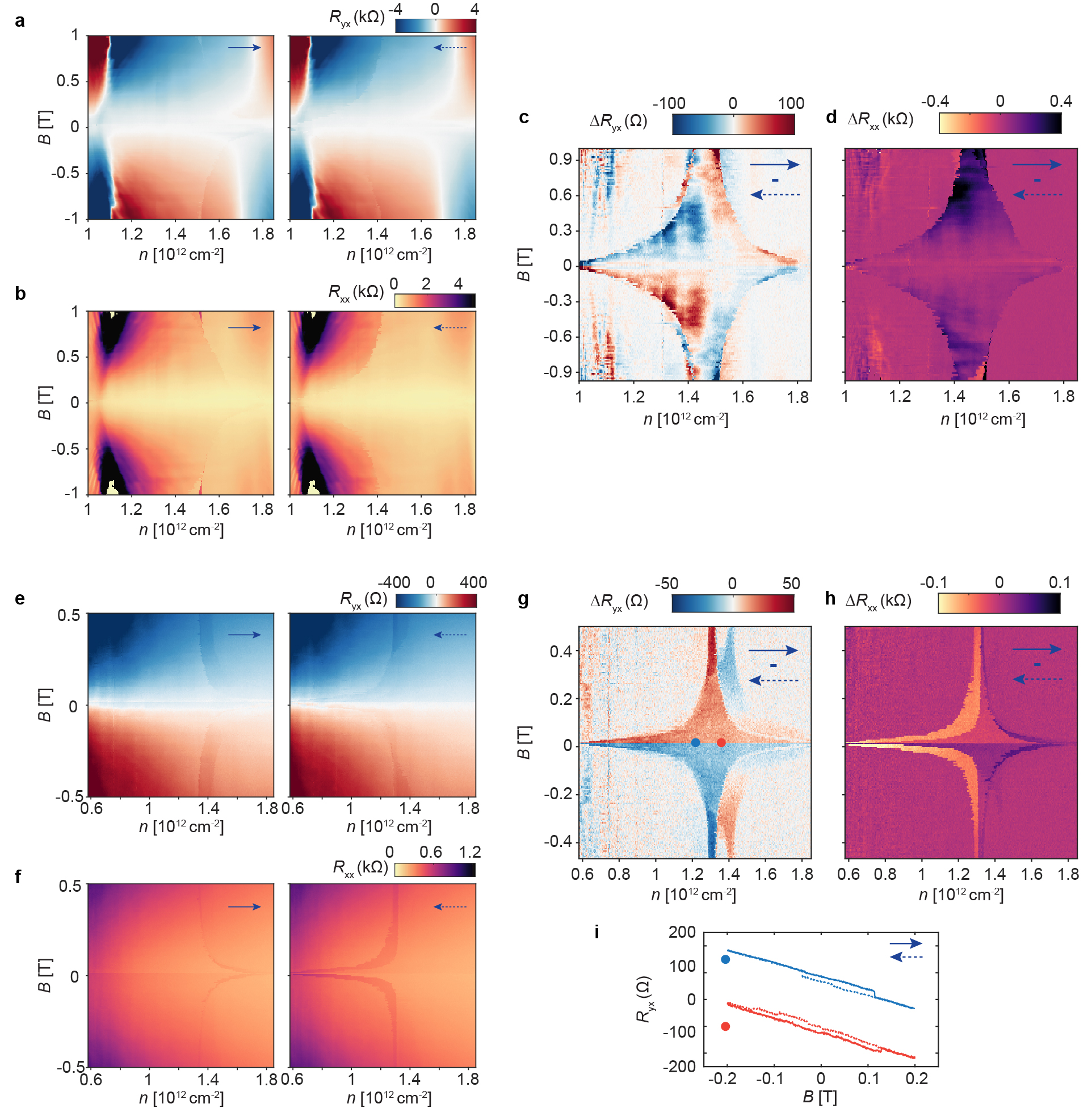}
    \caption{
    \textbf{Electrical switching behavior and magnetization reversal near $\nu$= 1 in devices D4 and D5.} 
    \textbf{a, b}, $R_{yx}$ and $R_{xx}$ as a function of $n$ and $B$ measured in device D4 with forward (left panels) and backward (right panels) sweeps of $n$. 
    \textbf{c, d}, The differences in $R_{xx}$ and $R_{yx}$, denoted  by $\Delta R_{yx}$  and $\Delta R_{xx}$, respectively, between forward and backward sweeps shown in (a) and (b). $\Delta R_{yx}$ was antisymmetrized with respect to the magnetic field.
    \textbf{e, f}, $R_{yx}$ and $R_{xx}$ as a function of \textit{n} and \textit{B}, measured in device D5 with forward (left panels) and backward (right panels) sweeps of $n$. 
     \textbf{g, h}, Differences $\Delta R_{yx}$ and $\Delta R_{xx}$, respectively, between forward and backward sweeps shown in (e) and (f). 
$\Delta R_{yx}$ was antisymmetrized with respect to the magnetic field.
     The trajectories along which the data in \textbf{a, b, e, f} were acquired are indicated by the dashed lines in Fig.~\ref{fig:S:DevicesResistance} \textbf{d, e}. 
     \textbf{i}, Magnetic hysteresis of $R_{yx}$ measured on both sides of the switching transition, as indicated in (g), reverses sign, confirming  the presence of magnetization reversal. 
    }
    \label{fig:S:D3andD4switching}
\end{figure}

\clearpage

\newpage\clearpage
\renewcommand{\figurename}{\textbf{Supplementary Fig.}}
\renewcommand{\thefigure}{S\arabic{figure}}
\setcounter{figure}{0}

\renewcommand{\tablename}{\textbf{Supplementary Tab.}}
\renewcommand{\thetable}{S\arabic{table}}
\setcounter{table}{0}
\setcounter{secnumdepth}{2}

\widetext
\section*{Supplementary Information}

\section{Details of the simulations}

\subsection{Continuum model}\label{sec:numerical_details}

The continuum model for tMBG is a generalization of the Bistritzer-MacDonald model~\cite{Bistritzer2011}. Without proximity to WSe$_2$, the model is independent of spin. In the $K$-valley, it can be written as 
\begin{equation}\label{eq:Hamiltonian}
H_K = \begin{pmatrix}
    H_0(\bm{k}_1) - \frac{1}{2}\Delta V & U & \\
    U^\dagger & H^A_0(\bm{k}_2) & g(\bm{k}_2) \\
    & g^\dagger(\bm{k}_2) & H^B_0(\bm{k}_2) + \frac{1}{2}\Delta V
\end{pmatrix}.
\end{equation}

Here, $\Delta V$ is the potential difference between the top and bottom layers of graphene. It is related to the displacement field $D$ by $\Delta V = \frac{2Dd}{\epsilon_{g,\perp}}$, where $d$ is the interlayer distance and $\epsilon_{g,\perp}$ is the dielectric constant of graphene in the perpendicular direction. Unless otherwise noted, we usually take $\epsilon_{g,\perp} = 4$, but we caution that its correct value is not known very well and hence should be viewed as a phenomenological parameter. In Eq.~\ref{eq:Hamiltonian}, $\bm{k}_l = \bm{k} - \bm{K}_l$ is the momentum measured from the (rotated) Dirac point of the monolayer graphene ($l = 1$) and the Bernal bilayer graphene ($l = 2$) respectively. We define $k_\theta = \frac{8\pi}{3\sqrt{3}a_{\text{CC}}}\sin(\theta/2)$ where $a_{\text{CC}} = 1.42 \times 10^{-10}\,$m is the length of C-C bond. The primitive vectors of the moir\'e reciprocal lattice are then given by $\bm{G}_1 = k_\theta(\sqrt{3}, 0)$ and $\bm{G}_2 = k_\theta(-\frac{\sqrt{3}}{2}, \frac{3}{2})$, and the Dirac points of the two layers are related by $\bm{K}_1 - \bm{K}_2 = \frac{1}{3}\bm{G}_1 + \frac{2}{3}\bm{G}_2$. The monolayer and bilayer parts of the Hamiltonian are given by
\begin{align}
    H_0 &= v\bm{k} \cdot \bm{\sigma} \\
H^A_0(\bm{k}) &= H_0(\bm{k}) + \frac{1}{2}(I_2 - \sigma^z)\Delta^\prime \\
    H^B_0(\bm{k}) &= H_0(\bm{k}) + \frac{1}{2}(I_2 + \sigma^z)\Delta^\prime \\
        g(\bm{k}) &= \begin{pmatrix}
        -\hbar v_4 k_- & -\hbar v_3 k_+ \\
       \gamma_1 & -\hbar v_4 k_-
    \end{pmatrix},
\end{align}
where $\bm{\sigma} = (\sigma^x, \sigma^y)$ are the Pauli matrices in sublattice space, $v$ is the graphene Fermi velocity, and $k_\pm = k_x \pm ik_y$.  $H_0$ is the Hamiltonian of monolayer graphene, $H^A_0$ and $H_0^B$ includes the correction to onsite energy due to Bernal stacking, and $g$ represents hopping between the two layers within each Bernal bilayer stack. The parameters are given by $\gamma_0 = 2.610$~eV, $\gamma_1 = 0.361$~eV, $\gamma_3 = 0.283$~eV, $\gamma_4 = 0.138$~eV and $\Delta^\prime = 0.015\,$eV~\cite{jung_accurate_2014}, where the $v_i$'s are further related to the $\gamma_i$'s by $v_i = \frac{3a}{2\hbar}\gamma_i$ and $v \equiv v_0$. The interlayer moir\'e coupling is the same as that of twisted bilayer graphene in the Bistritzer-MacDonald model, i.e.,
\begin{equation}
    U = T_1 + T_2e^{i(\bm{G}_1 + \bm{G}_2) \cdot \bm{r}} + T_3e^{i\bm{G}_2 \cdot \bm{r}},
\end{equation}
where
\begin{equation}
    T_{n + 1} = w_{\text{AA}} + w_{\text{AB}}(\sigma^x \cos(n\phi) + \sigma^y \sin(n\phi)),
\end{equation}
with $\phi = 2\pi/3$. We take $w_{AB} = 0.11$~eV, $w_{AA} = 0.08$~eV, unless otherwise specified. The Hamiltonian for valley $K'$ is given by time-reversal symmetry.

\subsection{Ising spin-orbit coupling}

Proximity to WSe$_2$ generates spin-orbit coupling (SOC). In the main text, we focus on Ising SOC, which is given by
\begin{equation}
      H_{\text{Ising}} = \frac{\lambda_I}{2}\tau_z s_z, 
\end{equation}
on the graphene layer immediately adjacent to the WSe$_2$ (i.e., the third column and row in Eq.~\ref{eq:Hamiltonian}).
Below we discuss how Rashba SOC can be modeled and provide quantitative evidence that it is safe to ignore it.

\subsection{Interaction and Hartree-Fock}

In order to calculate the orbital magnetization, we first obtain the interaction-renormalized band structure. The interaction Hamiltonian is given by
\begin{equation}\label{eq:INT}
    H_{\text{int}} = \frac{1}{2A}\sum_{\bm{q}}V(\bm{q}):\rho_{\bm{q}}\rho_{-\bm{q}}:
\end{equation}
where $\rho_{\bm{q}}$ is the band-projected density operator given by
\begin{equation}\rho_{\bm{q}} = \sum_{\bm{k} \in \text{BZ}, ab\tau s} \braket{u_{\bm{k}+\bm{q},\tau a}|u_{\bm{k},\tau b}}c^\dagger_{\bm{k} + \bm{q},\tau s a} c^{\phantom\dagger}_{\bm{k},\tau s b},
\end{equation}
where $a,b$ are band indices, $\tau$ is the valley index and $s$ is the spin index. We consider the dual-gate screened Coulomb potential $V(q)=e^2\tanh(qd)/(2\epsilon_0\epsilon_r q)$, with gate screening distance $d=25$\,nm and dielectric constant $\epsilon_r = 20$. To avoid double counting of interaction, we use the ``average" subtraction scheme, where the reference density corresponds to half-filling of the central bands, fully filling remote valence bands and leaving remote conduction bands empty. This interaction was then incorporated through a self-consistent Hartree–Fock mean-field approach.

\subsection{Orbital magnetization calculations}

For a Bloch Hamiltonian $H(\bm{k})$ with Bloch functions $n(\bm{k})$ and energy $\epsilon_n(\bm{k})$, the orbital magnetization is given by~\cite{xiao_berry_2005,thonhauser_orbital_2005,ceresoli_orbital_2006,shi_quantum_2007}
\begin{equation}
    M_z=\frac{e}{\hbar}\mathrm{Im}\int\frac{d\bm{k}}{(2\pi)^2}\sum_n\left\langle\partial_{k_x}n\middle|H(\bm{k})+\epsilon_n(\bm{k})-2\mu\middle|\partial_{k_y}n\right\rangle\theta\left(\mu-\epsilon_n(\bm{k})\right),
\end{equation}
where $\mu$ is the chemical potential and $\theta$ is the Heaviside function. The orbital magnetization can be decomposed into two components
\begin{align}
m_n^{(1)}(\bm{k})
&=\frac{e}{\hbar}\mathrm{Im}\left\langle\partial_{k_x}n\middle|H(\bm{k})-\epsilon_n(\bm{k})\middle|\partial_{k_y}n\right\rangle\\
&=-\frac{e}{\hbar}\mathrm{Im}\sum_{m\neq n}
\frac{
\left\langle n\middle|\partial_{k_x}H(\bm{k})\middle|m\right\rangle
\left\langle m\middle|\partial_{k_y}H(\bm{k})\middle|n\right\rangle
}{
\epsilon_n(\bm{k})-\epsilon_m(\bm{k})
}
\end{align}

\begin{align}
m_n^{(2)}(\bm{k})
&=-\frac{2e}{\hbar}\mathrm{Im}\left\langle\partial_{k_x}n\middle|\mu-\epsilon_n(\bm{k})\middle|\partial_{k_y}n\right\rangle\\
&=-\frac{2e}{\hbar}\mathrm{Im}\sum_{m\neq n}
\frac{
\left\langle n\middle|\partial_{k_x}H(\bm{k})\middle|m\right\rangle
\left\langle m\middle|\partial_{k_y}H(\bm{k})\middle|n\right\rangle
}{
\left(\epsilon_n(\bm{k})-\epsilon_m(\bm{k})\right)^2
}
\left[\mu-\epsilon_n(\bm{k})\right]\\
&=\frac{e}{\hbar}\Omega_n(\bm{k})\left(\mu-\epsilon_n(\bm{k})\right),
\end{align}
where $\Omega_n(\bm{k})$ is the Berry curvature.

In order to compute the orbital magnetization of the interacting model, we perform self-consistent Hartree-Fock to obtain the renormalized band structure, which is then used as input into the above expressions to calculate the orbital magnetization. One difficulty in this approach is that it is necessary to include a large number of remote bands in order to obtain reliable estimates of orbital magnetization, but performing self-consistent Hartree-Fock with such a large number of remote bands is too computationally costly to be practical. As such, we adopt a procedure similar to that used in Ref.~\cite{xie_unconventional_2025}, wherein we first perform self-consistent Hartree-Fock with a small number of bands (6 bands per spin and valley, unless otherwise specified), and then embed the result into a larger Hilbert space with more remote bands (48 bands per spin and valley sector, unless otherwise specified).

\section{Effect of the Ising and Rashba SOC on spin polarization}\label{sec:rashba}

\begin{figure}[ht!]
    \centering
    \includegraphics[width=\linewidth]{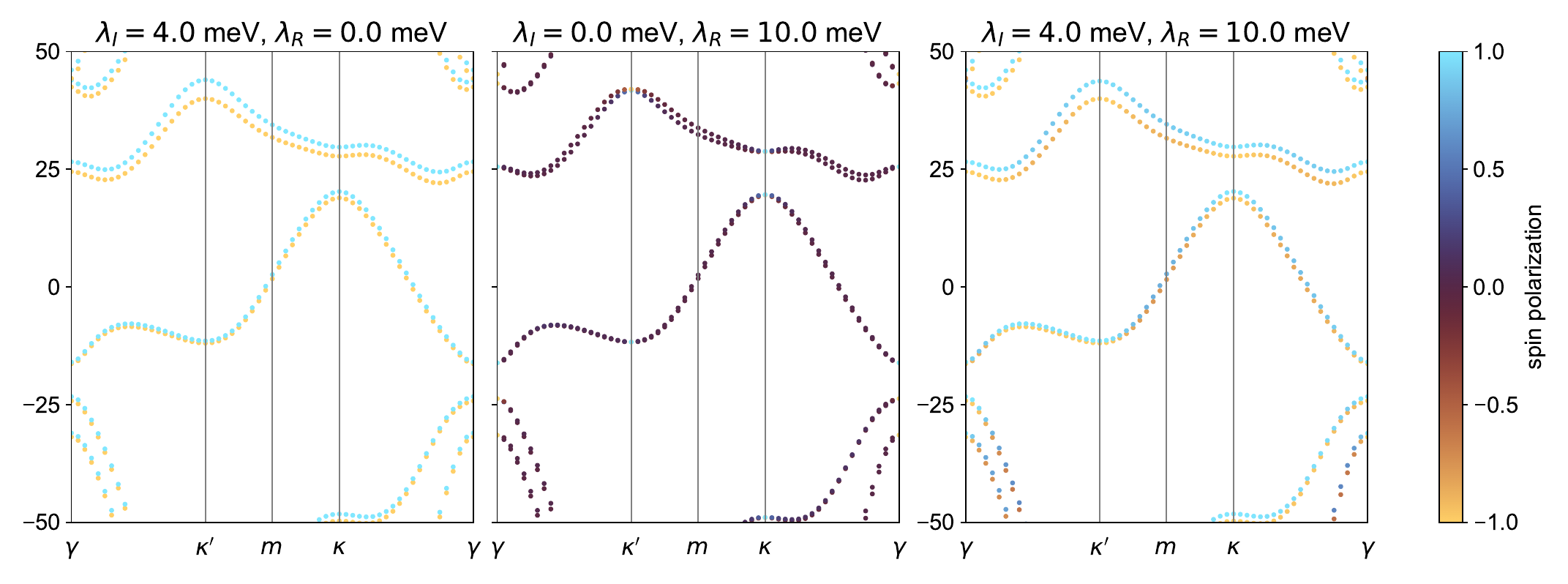}
    \caption{Band structure of tMBG with $D = 0.5$~V/nm and $\theta = 1.23^\circ$ with different combinations of Ising and Rashba SOC. We observe that the band structure with $\lambda_I = 4.0$~meV and $\lambda_R = 10.0$~meV is almost indistinguishable from the band structure with only $\lambda_I = 4.0$~meV.}
    \label{fig:rashba_vs_Ising}
\end{figure}

Besides Ising SOC, proximity to WSe$_2$ can also induce Rashba SOC. Together, they can be written as 
\begin{equation}
    H_{\text{SOC}} = \frac{\lambda_I}{2}\tau_z s_z + \frac{\lambda_R}{2}(\tau_z \sigma_x s_y - \sigma_y s_x),
\end{equation}
where $\tau_i, \sigma_i, s_i$ denote Pauli matrices in valley, sub-lattice and spin degrees of freedom respectively, and $\lambda_R$ parametrizes the strength of the Rashba term. Even though $\lambda_R$ may be of similar or even larger magnitude as compared to $\lambda_I$, as shown in Fig.~\ref{fig:rashba_vs_Ising}, the effect on band structure due to $\lambda_R$ is much smaller than that of $\lambda_I$, and when both are present, the band structure is very similar to that of Ising SOC alone.
Importantly for our interpretation of the influence on MR-driven switching, even when both Ising and substantial Rashba spin-orbit coupling are present, the conduction band at $D>0$ still has predominantly out-of-plane spin polarization which is locked to valley. 

\section{Explanation of the magnetization reversal at $1<\nu<2$}

In this discussion, we maintain the following convention: the conduction band of the $K$-valley, under positive displacement field $D$, has a \textit{positive} Chern number,i.e, $+2$. The sign convention for magnetization is related to the Chern number, namely for an insulator with a positive Chern number, the orbital magnetization on the $n$-doped side is more \textit{positive} than the $p$-doped side. The labeling of the spin is defined such that $\bm{\uparrow}$ spin contributes to positive magnetization. We consider magnetic field in a direction that energetically favors positive (spin and orbital) magnetization, e.g. $\uparrow$ spin has lower energy than $\downarrow$ spin.

With the sign conventions fixed, we make a non-trivial assumption that Ising SOC lowers the energy of $K \downarrow$ and $K' \uparrow$ relative to $K \uparrow$ and $K' \downarrow$. If we write the SOC term as $\frac{\lambda_I}{2} \tau_z s_z$, we are setting $\lambda_I > 0$. Wherever relevant, we also assume that magnetic field is sufficiently small that the energetic contributions due to (spin and orbital) magnetization in the magnetic field are always smaller than the SOC.

As a preliminary, let us first examine the energy competition when no SOC is applied. For $\nu = 1$, we occupy 1 out of 4 spin-valley sectors. Clearly, regardless of the sign of orbital magnetization, it is energetically preferable to fill the $\uparrow$ spin sectors than the $\downarrow$ spin sectors. Therefore, the competition is between $K\uparrow$ and $K'\uparrow$. For $\nu = 3$, we always occupy both valleys of the $\uparrow$ sector, and the competition is between $K \downarrow$ and $K' \downarrow$. In both cases, we observe that the competition is between two valleys with the same spin. As such, magnetic reversal corresponds to a reversal in the sign of \textit{orbital} magnetization.

This picture needs to be modified when SOC is applied. As mentioned above, we work in the small field limit where SOC dominates over magnetization energy. For $\nu = 1$, the two competing sectors are $K \downarrow$ and $K' \uparrow$. For $\nu = 3$ however, since both $K \downarrow$ and $K' \uparrow$ are filled, the competition is between filling $K \uparrow$ and $K' \downarrow$. We observe:

\begin{itemize}
    \item When SOC is included, for both $\nu = 1$ and $\nu = 3$, the competition is now between sectors with opposite spins. As such, we need to compare the total magnetization, instead of just the orbital magnetization.
    \item The spins contribute in opposite directions for $\nu = 1$ and $\nu = 3$.
\end{itemize}

Now, more specifically, let us focus on the metallic filling $1 < \nu < 2$. As the Ising SOC favors $K \downarrow$ and $K' \uparrow$ relative to $K \uparrow$ and $K' \downarrow$, electrons occupy sectors $K \downarrow$ and $K' \uparrow$. Due to strong tendency towards flavor polarization, one of the two sectors has 1 electron per unit cell (i.e., the conduction band is fully occupied), and the other sector has $\nu - 1$ electron per unit cell (this is verified numerically). Out of the two ferromagnetic metallic states related by time-reversal symmetry, we call the state with fully occupied $K \downarrow$ and partially occupied $K' \uparrow$ sector as the $K$-polarized state, with orbital magnetization $M^K_{\mathrm{orbital}}(\nu)$. The spin magnetization of the $K$-polarized state has magnitude $|M_{\mathrm{spin}}(\nu)| = 2 - \nu$, measured in Bohr magneton per unit cell, and has a \textit{negative} sign. The total magnetization of the $K$-polarized state is therefore $M^K_{\text{tot}}(\nu) \equiv M^K_{\mathrm{orbital}}(\nu) + M_{\mathrm{spin}}(\nu) = M^K_{\mathrm{orbital}}(\nu) - |M_{\mathrm{spin}}(\nu)|$. Magnetic reversal can happen if, from $\nu = 1$, one starts off with positive total magnetization, so the $K$-polarized state is the preferred ground state under a small magnetic field, and $M^K_{\mathrm{orbital}}(\nu) - |M_{\mathrm{spin}}(\nu)|$ changes sign at some metallic filling to flip the ground state from $K$-polarized to $K'$-polarized. 

\begin{figure}
    \centering
    \includegraphics[width=\linewidth]{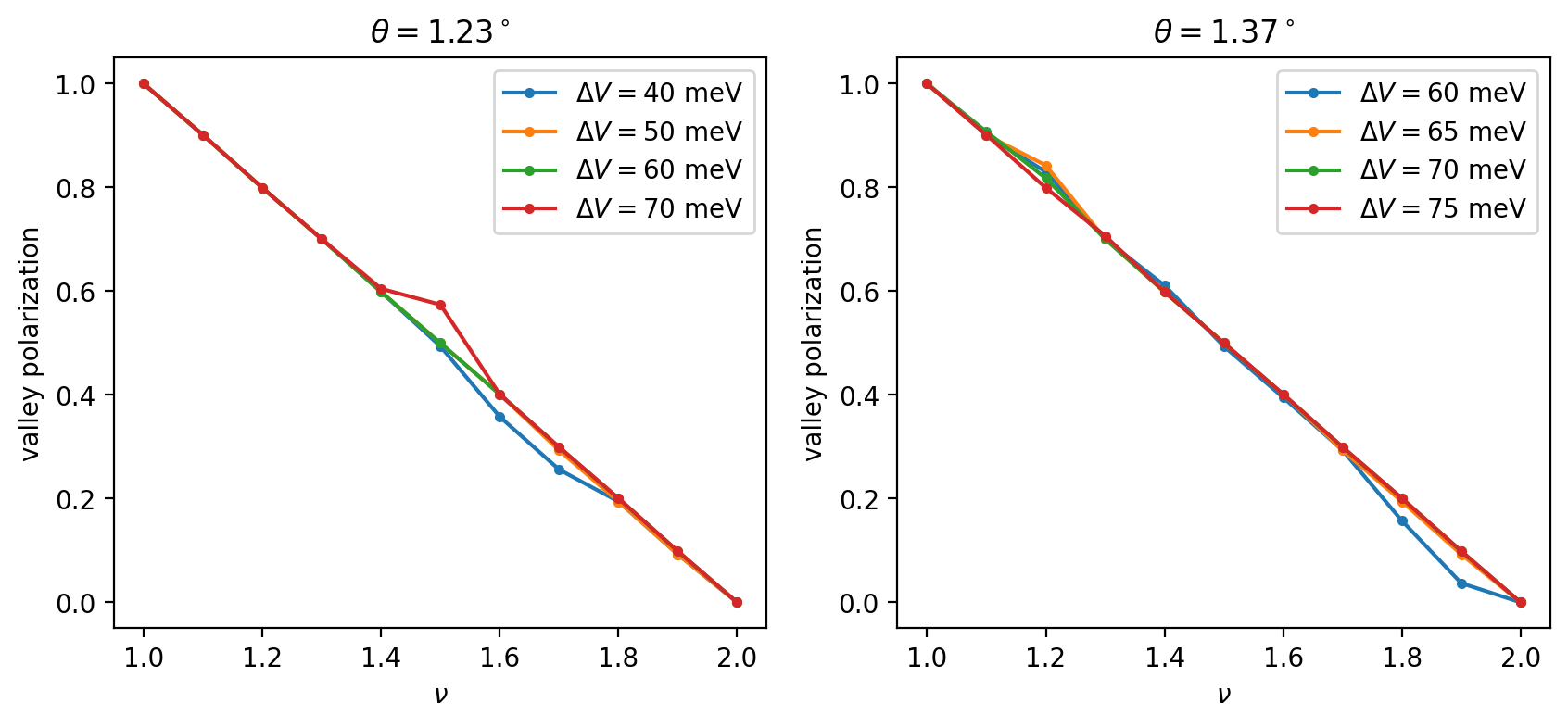}
    \caption{Valley polarization obtained from $18 \times 18$ self-consistent Hartree-Fock for $1 < \nu < 2$. We observe that, to a very good extent, the valley polarization varies linearly with $\nu$. We scan over a range of $\Delta V$ to account the uncertainty in the conversion between displacement field $D$ and $\Delta V$.}
    \label{fig:vpol}
\end{figure}

\begin{figure}
    \centering
    \includegraphics[width=\linewidth]{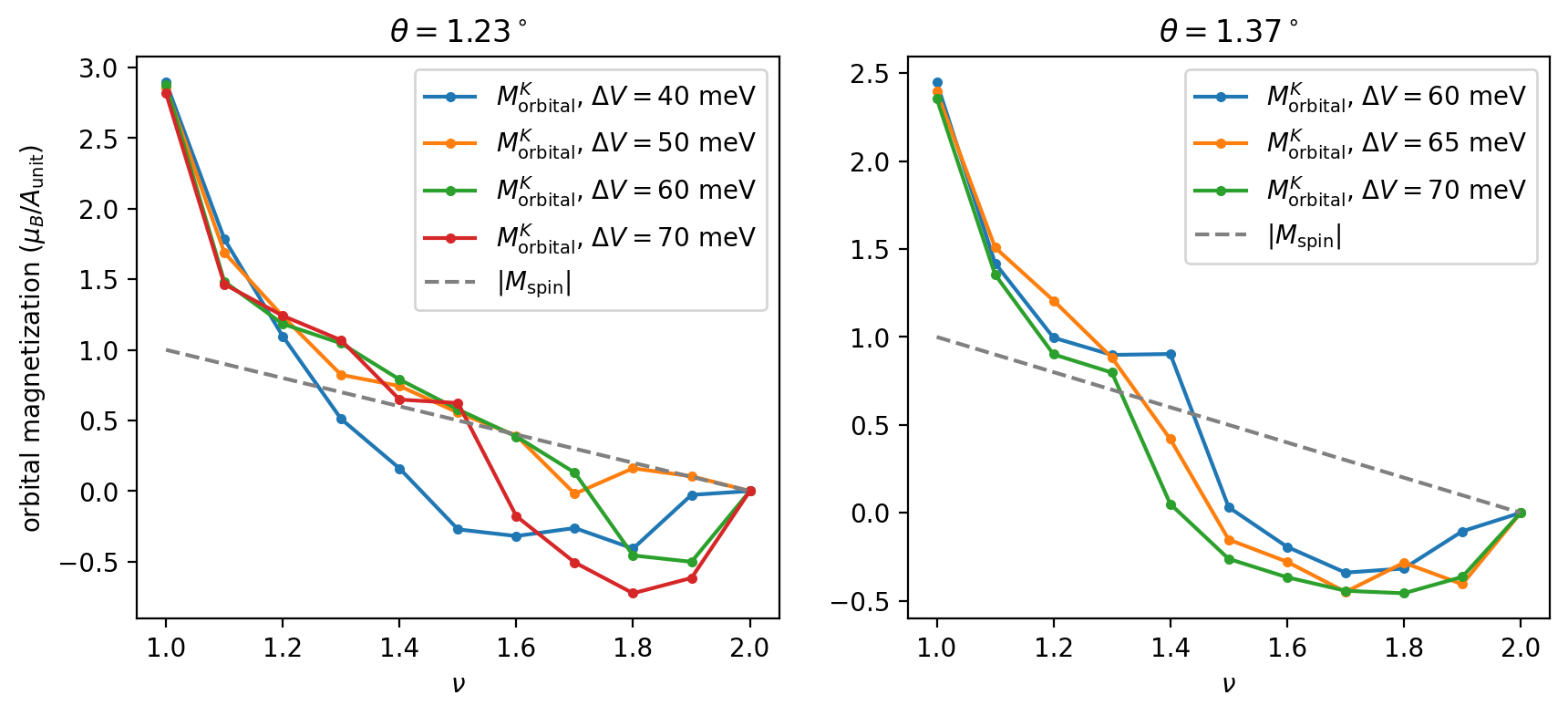}
    \caption{Orbital magnetization as a function of filling factors. The orbital magnetization is obtained from $18 \times 18$ Hartree-Fock. We observe that, across a variety of parameters, $M^K_{\mathrm{orbital}}(\nu) - |M_{\mathrm{spin}}(\nu)|$ changes sign at some metallic filling.  We scan over a range of $\Delta V$ to account the uncertainty in the conversion between displacement field $D$ and $\Delta V$.}
    \label{fig:magnetization_vs_filling}
\end{figure}

Numerically, we first verify that at metallic filling factors, one sector has 1 electron per unit cell and the other sector has $\nu - 1$ electron per unit cell. To do so, we perform self-consistent Hartree-Fock at metallic fillings, and calculate the valley polarization. As shown in Fig.~\ref{fig:vpol}, to a very good extent, the valley polarization varies linearly with the filling factor, consistent with our assumption. Then, we compute the orbital magnetization using the method outlined 
above. 
As shown in Fig.~\ref{fig:magnetization_vs_filling}, for a variety of parameters (twist angle $\theta$ and interlayer potential $\Delta V$, which is proportional to the displacement field), $M^K_{\mathrm{orbital}}(\nu) - |M_{\mathrm{spin}}(\nu)|$ changes sign at some metallic filling, consistent with the observation of magnetic reversal. Here, we have scanned over a range of $\Delta V$ to account the uncertainty in the conversion between displacement field $D$ and $\Delta V$.

\end{document}